\documentclass{aa} 

\usepackage{natbib}
\bibpunct{(}{)}{;}{a}{}{,}
\usepackage{graphicx}
\usepackage{lscape}
\usepackage{aalongtable}
\usepackage{subfigure}
\usepackage{txfonts}
\newcommand{\Ha}{\mbox{H$\alpha$}}
\newcommand{\Hb}{\mbox{H$\beta$}}
\newcommand{\Hd}{\mbox{H$\delta$}}
\newcommand{\Hg}{\mbox{H$\gamma$}}

\begin{document}
\title{Mapping the starburst in Blue Compact Dwarf Galaxies
}

\subtitle{PMAS Integral Field Spectroscopy of Mrk~1418\thanks{Based on 
observations obtained at the German-Spanish Astronomical Center,
Calar Alto, operated by the Max-Planck-Institut fur Astronomie Heidelberg
jointly with the Spanish National Commission for Astronomy.}}

\author{L. M. Cair{\'o}s
         \inst{1}
         \and
        N. Caon
         \inst{2}
         \and
        C. Zurita
         \inst{2}
         \and
        C. Kehrig
         \inst{3}
         \and
        P. Weilbacher
         \inst{1}
         \and
        M. Roth
         \inst{1}
}
       
\institute{Astrophysikalisches Institut Potsdam,
           An der Sternwarte 16, D-14482 Potsdam, Germany\\
           \email{luzma;pweilbacher;mmroth@aip.de}
           \and
           Instituto de Astrof{\'\i}sica de Canarias, E-38200 La Laguna, 
           Tenerife, Canary Islands, Spain\\ 
           \email{nicola.caon;czurita@iac.es}
           \and
           University of Michigan Department of Astronomy
           830 Dennison Bldg,
           500 Church St,
           Ann Arbor, MI 48109-1042 \\
           \email{kehrig@umich.edu}
} 

\date{}

\abstract
{}
{By means of optical Integral Field Spectroscopy observations, we aim to
disentangle and characterize the starburst component in the Blue Compact
Dwarf Galaxy Mrk~1418. In particular we propose to study the stellar and
ionized gas morphology, to investigate the  ionization mechanism(s) acting in
the interstellar medium, to derive the physical parameters and abundances of
the ionized gas.}
%
{Integral Field Spectroscopy observations of Mrk~1418 were carried out with
the Potsdam  Multi-Aperture Spectrophotometer (PMAS) at the 3.5 m telescope 
at Calar Alto Observatory. The central $16\arcsec\times 16\arcsec$ 
($1.14\times1.14$ kpc$^{2}$ at the distance of Mrk~1418) were mapped with a 
spatial sampling of $1\arcsec$; we took data in the 3590-6996 \AA\ spectral 
range, with a linear dispersion  of 3.2 \AA\ per pixel. 
The seeing was about $1\farcs5$.
From these data we built maps of the most
prominent emission lines, namely [\ion{O}{ii}], \Hb, [\ion{O}{iii}], \Ha,
[\ion{N}{ii}] and  [\ion{S}{ii}] as well as of several continuum bands,
plus maps of the main line ratios: [\ion{O}{iii}]/\Hb,
[\ion{N}{ii}]/\Ha, [\ion{S}{ii}]/\Ha, and \Ha/\Hb, and derived the
physical parameters and gaseous metal abundances of the different
star-forming regions detected in the field of view.}
%
{Mrk~1418 shows a distorted morphology both in the continuum and in the
ionized gas maps; the current star- formation episode is taking place in five
knots, distributed around the nucleus of the galaxy. The interstellar medium
surrounding  these knots is photo-ionized by stars, with no clear evidence for
other excitation mechanisms. The galaxy displays an inhomogeneous dust 
distribution, with the high \Ha/\Hb\ ratio in the central areas indicating  
a large amount of dust. The oxygen abundances derived for the individual  
star-forming knots are very similar, suggesting that the ionized interstellar 
medium is chemically homogeneous in O/H over spatial scales of hundreds of 
parsecs.
This abundance ($Z\approx0.4Z_{\sun}$ from the empirical calibrations) 
places Mrk~1418 among the high metallicity BCD group.} 
%
{
These findings show the advantages of IFS when investigating the 
properties of such complex objects as BCDs, with an asymmetric star forming 
component. Only a bidimensional mapping of their central regions allows to
approach such questions as the star formation processes in BCDs, the 
star-forming history of the individual starburst knots, or the abundance 
gradients.
}

\keywords{galaxies:dwarf -- galaxies:starburst -- galaxies: stellar populations
--galaxies:individual (Mrk1418) --ISM: HII regions}
\maketitle

%

\section{Introduction}

Blue Compact Dwarf (BCD) galaxies have a unique potential to address
many central issues in contemporary galaxy research. Chemically
unevolved nearby star-forming (SF) dwarfs, such as BCDs, are in fact
an important link to the early Universe and the epoch of galaxy
formation, as they are regarded as the local counterparts of the
distant subgalactic units (building blocks) from which larger systems
are created at high redshifts
\citep{Kauffmann1993,Lowenthal1997}; the study of these systems may
provide important insights into the star-formation process of
distant galaxies. Moreover, even though most BCDs are not 
genuinely young galaxies, their metal deficiency make them useful objects to 
set constraints on the primordial $^{4}$He abundance, and to monitor the
synthesis and dispersal of heavy elements in a nearly pristine
environment \citep{Pagel1992,Masegosa1994,Izotov1997,Kunth2000}. BCDs
are also ideal laboratories for the study of the starburst phenomenon:
being smaller and less massive than normal galaxies, they cannot
sustain a spiral density wave and do not suffer from disk
instabilities, which considerably simplifies the study of the
star-formation process. Besides, the radiation emitted by their SF
regions is less affected by the stellar continuum than in giant spiral
galaxies, allowing for more accurate studies of element abundances.

In spite of the great effort spent on the study of BCDs during the last two
decades, many fundamental questions such as the
mechanisms responsible for the ignition of the starburst, their evolutionary 
status, or their star formation histories, are still far from being well
understood. Detailed spectrophotometric analyses of individual nearby BCDs,
providing the information needed to disentangle young and old stellar
populations, as well as the
gas and the dust components, are crucial to further investigate these
topics \citep{Papaderos1998,GildePaz2000,Cairos2002,Guseva2003a,Guseva2003b,
Guseva2003c,Cairos2007}.

Yet this kind of analysis is scarcely found in the literature \citep[see][and
references therein]{Cairos2007}. This fact is probably due to the large amount
of time needed to obtain the required data using traditional observing 
techniques, i.e. the combination of broad and narrow-band imaging and
long-slit spectroscopy; this can easily lead to observing times of several
nights per object. Moreover, these observations typically suffer from varying 
atmospherics conditions, which can render the combination of the various  data
sets a complicated and tedious undertaking.

The relatively new observational method of Integral Field Spectroscopy (IFS)
allows to overcome these problems. Each single exposure with an Integral Field
Unit (IFU) provides at the same time both spatial and spectral information,
making IFU observations an order of magnitude more efficient than traditional
observing techniques. Furthermore, IFS provides simultaneous spectra for all
spatial resolution elements, under the same instrumental and atmospheric
conditions, resulting in a homogeneous dataset. As several recent
studies have shown \citep{Izotov2006,GarciaLorenzo2008,Kehrig2008,Vanzi2008}
Integral Field Spectrographs are ideal instruments to study such small and yet
complex objects as BCDs.

We have undertaken a comprehensive spectrophotometric analysis of a large and
representative sample of BCDs by means of IFS. This sample, which includes
$\approx$ 40 objects, covers the whole range in luminosity (from $M_B
= -21$ to $M_B = -14$) and morphologies (nE, iE, IC, and iI,
following the classification scheme introduced by \cite{Loose1986}) that we 
find among BCD galaxies. 

The dataset resulting from this study will help answer some of the most
crucial open questions in BCD research: to effectively disentangle the young
and old stellar populations, to set constraints on the age of the galaxies, to
study the kinematics of the gas and stars, to look for signatures of
merger or interaction episodes, and, finally, to determine their star-forming
histories. Besides, these data will provide the essential template to
understand and interpret the results for intermediate and high-z star-forming
galaxies.

This is the first of a series of papers in which the outcomes from  this
project will be presented. In this paper we carry out a pilot study of the
galaxy \object{Mrk 1418}, with the intention of showing the great potential of
IFS when applied to the BCDs field. \object{Mrk 1418} is part of a sample of
ten objects which were observed with the Potsdam Multi-Aperture
Spectrophotometer (PMAS) at the 3.5 m telescope at Calar Alto; results for the
remaining galaxies observed with PMAS will be presented in a forthcoming 
paper.

\object{Mrk 1418} is a low luminosity ($M_{B} = -16.96$) BCD which, according 
to the morphological classification by \cite{Loose1986}, belongs to the most 
common BCD class, the iE BCDs. 
It has been included in the BCDs samples
analyzed by \cite{Doublier1997,Doublier1999}, \cite{GildePaz2003} and
\cite{GildePaz2005} as well as in the recently published ``Atlas of Markarian
galaxies'' by \cite{Petrosian2007}. Its basic data are shown in Table
\ref{Table:datamrk1418}.
Its proximity (distance of 14.6 Mpc) and the
asymmetric, clumpy starburst morphology make this object an excellent
test-bench to illustrate the need of bidimensional spectroscopy to
analyze BCD galaxies.
Its current SF activity takes place in a number of knots placed on a regular
and elliptical envelope ($B-R \approx 1.6$;
\citealp{Doublier1997,GildePaz2005}). \cite{VanZee2001} found, from high
spatial resolution \ion{H}{i} synthesis observations, that the neutral gas in 
Mrk~1418 extends to approximately two times the optical diameter of the galaxy
($D_{25}=48$ arcsec, $D_\textrm{HI}/D_{25}=2.3$, see Table 1 in 
\citealp{VanZee2001}) 
and displays kinematical peculiarities reminiscent of tidal tails. 
Although the optical profile is better fitted by a de Vaucouleurs law 
\citep{Doublier1997},
the HI radio observations presented in \cite{VanZee2001}
indicate that the galaxy is supported by rotation.

The paper is structured as follows: in Sect.~\ref{Sect:Data} we describe the
observations, the data reduction process and the method employed to derive the
two-dimensional maps. In Sect.~\ref{Sect:Results} we present the main results
of this work, that is, the emission line and continuum flux, emission line
ratio, and velocity maps, as well as the results derived from the analysis of
the integrated spectra of the selected galaxy regions. These results are
discussed and summarized in Sect.~\ref{Sect:Discussion} and
Sect.~\ref{Sect:Conclusions}, respectively.

\section{The data}
\label{Sect:Data}

\subsection{Observations}
\label{SubSect:Observations}

Mrk~1418 was observed in 2007 March with the Potsdam Multi-Aperture
Spectrophotometer (PMAS) at the 3.5 m telescope at the Observatorio
Astron{\'o}mico Hispano Alem{\'a}n, Calar Alto (CAHA). PMAS is an integral
field spectrograph, with a lens array of $ 16\times 16$ square elements,
connected to a bundle of fiber optics, whose 256 fibers are re-arranged to
form a pseudoslit in the focal plane of the spectrograph (see
\citealp{Roth2005} and \citealp{Kelz2006} for more details about the
instrument). In the configuration we used, 
each lenslet covers  $1\arcsec\times1\arcsec$  in the sky, 
thus providing a total field of view of $16\arcsec\times16\arcsec$
($1.14\times1.14$ kpc$^{2}$ at Mrk~1418 distance, with a spatial 
sampling of 71~pc/$\arcsec$) . 

A grating with 300 grooves per mm was used during the observations, in
combination with a $2048 \times 4096$ pixel SITe ST002A CCD detector. 
This set-up provides a spectral range of 3590-6996 \AA, with a linear 
dispersion of 3.2 \AA\ per pixel (the CCD was binned $2\times2$ in the spatial 
and dispersion directions).
The actual spectral resolution, obtained by measuring the width of 
the arc lines in comparison spectra, was about 6.8 \AA\ FWHM, corresponding to
$\sim 300$ km s$^{-1}$ at \Ha\ and $\sim 400$ km s$^{-1}$ at \ion{O}{iii}.

We observed a total of 4800 s on the galaxy (four exposures of 1200 s each); 
additional sky frames were obtained by offsetting the telescope several
arcmin away from the target. Calibration frames were taken before
and after the galaxy exposures. These calibrations consist of spectra of
emission line lamps (HgNe lamp), which are required to perform the 
wavelength calibration, and spectra of a continuum lamp, necessary to 
locate the individual spectra on the CCD and to perform the flat-fielding
correction. As usual, bias and sky-flats exposures were taken at the
beginning and at the end of the night. The spectrophotometric standard
stars BD+75325 and BD+332642 were observed for flux calibration.
The seeing, as indicated by the CAHA seeing monitor, was about 
$1\farcs5$ FWHM.

\subsection{Data reduction}
\label{SubSect:DataReduction}

The data have been processed using standard IRAF\footnote{IRAF is distributed
by the National Optical Astronomy Observatories, which are operated by the
Association of Universities for Research in Astronomy, Inc., under cooperative
agreement with the National Science Foundation.} tasks.  The reduction
procedure includes bias subtraction and image trimming, tracing and
extraction, wavelength and distortion calibration, flat-fielding, combination
of the individual frames, sky-subtraction and flux calibration. 

The first step in the data reduction was the bias subtraction.  All the bias
exposures were averaged to obtain a master bias, which was then subtracted
from the rest of the frames.

Next, apertures (that is, the spectra produced by each fiber) were defined
on well exposed continuum frames using the task \textsc{apall}; the task first
finds the centroid of the aperture at a fixed position on the dispersion axis,
and then asks for the width of the  extraction window, which we set at 6.4
pixels (the best compromise between maximizing the flux from each spectrum and
avoiding contamination by nearby fibers). The apertures were then traced 
by fitting a polynomial to the centroid along the dispersion axis (because
of the field distortion introduced by the optics of the system, the apertures
show a noticeable curvature, depending on their spatial position). A fifth
degree Legendre polynomial was found to provide good fits, with a typical RMS
of about 0.01 pixels.  

Once the apertures were defined and traced in the continuum frames, we used
\textsc{apall} again to extract them in all the images. The extraction
consists on summing the pixels along the spatial direction into a final
one-dimensional spectrum. The final product is the so-called ``collapse''
spectrum: an image $M\times N$, where $M$ is the number of pixels in the 
dispersion direction, and $N$ is the number of spaxels (256 for PMAS).

Afterward we performed the wavelength calibration and the dispersion
correction. In order to calibrate in wavelength we used the tasks 
\textsc{identify} and \textsc{reidentify}: i) first, in the comparison spectra
(arc) we identified several emission features of known wavelength in a
reference fiber; ii) second, a fifth degree polynomial was fitted across the 
dispersion direction, 
resulting in a RMS of about one hundredth of \AA;  iii)
next, with  \textsc{reidentify} we identified the emission lines in the
remaining fibers of the arc frame, using the selected one as reference. The
polynomial fit derived in this way was used to calibrate the rest of 
the images (galaxy, standard stars and sky frames) using the task
\textsc{dispcor}.

The data were corrected for throughput (that is, spaxel-to-spaxel overall
sensitivity variations) and response (detector pixel non-uniformity) using
the task \textsc{msresp1d}. The individual galaxy spectra were then corrected
for atmospheric extinction, adopting the ``summer extinction coefficients''
published by \cite{Sanchezetal2007},  and combined using the task 
\textsc{imcombine}.

The next step was the sky subtraction. Sky frames (5 minutes
exposures) were processed in the same way as galaxy spectra. A one-dimensional
sky spectrum was produced by averaging, with a sigma-clipping algorithm, the
signal along the spatial direction. The flux of the three to four brightest
sky lines was measured in both the final sky spectrum and in the final galaxy
spectrum to determine the scale factor by which to multiply the sky spectrum
before subtracting it from the galaxy spectrum. Because the relative intensity
of different sky lines vary noticeably on short time scales \citep{Patat2003},
it is very difficult to find a scaling factor that applies equally well to all
the sky lines and the sky continuum, and some fine tuning is required.  The
final scale factor was found by trial and error, aiming at minimizing 
overall residuals in the sky lines (especially those close to galaxy
emission  or absorption lines) even if it leaves large residuals in bright sky
lines not affecting any interesting spectral features.

Finally, the flux calibration was performed. The spectra of the
two spectrophotometric standards observed the same night as Mrk~1418
were reduced in the same way as the galaxy frame,
except that the sky spectrum to be subtracted out was computed by 
median-combining the outermost fibers.
The one-dimensional spectra of the standard stars were obtained by 
summing all the fibers within a radius of about 2 FWHM (typically 3 to 4 
arcseconds) from the fiber with highest signal. The tasks \textsc{standard} 
and \textsc{sensfunc} were used to derive the sensitivity curves, after 
combining together the data for the different spectrophotometric stars 
observed in the same night.

By comparing the sensitivity curves obtained for all the spectrophotometric s
tandards observed throughout the three nights of the observing run, we
estimate that the relative uncertainty on the calibration factor is generally
equal or less than 2\%, except blueward of 4000 \AA, where the curves show a
marked change of slope and the uncertainty increases up to about 8\%.

\subsection{Emission Line Fits}
\label{SubSect:LineFit}

In order to measure the relevant parameters of the emission lines (center,
flux and width), they were fitted by a single Gaussian. The fit was carried
out by using the $\chi^{2}$ minimization algorithm implemented by C.~B.
Markwardt in the {\em mpfitexpr} IDL library\footnote{
http://cow.physics.wisc.edu/~craigm/idl/idl.html}. 

The continuum (typically 30-50 \AA\ on both sides) was fitted by a straight
line. Lines in a doublet were fitted forcing them to have the same redshift 
and width.

Criteria such as flux, error on flux, velocity, and width were used to do an
automatic assessment of whether to accept or reject a fit. For
instance, lines with too small ($< 2.5$ \AA) or too large ($> 5.0$ \AA)
widths were flagged as rejected, as well as lines with relative flux errors 
$\ga 10$\% --- the exact limits depending on the specific line. 
Such criteria were complemented by a visual inspection of all fits, which led 
to override, in a few cases (typically 5-10 spaxels), the automated 
criteria decision. For instance, sometimes the program fitted large 
spikes for very faint of even completely absent emission lines; in other 
cases, a particularly noisy continuum or bad pixels decreased the computed 
$S/N$ below the acceptance threshold, while the actual fit was clearly good 
enough.

For the H$\beta$ line, where absorption wings were present, we 
simultaneously fitted two gaussians, representing the absorption and the 
emission components.
To improve the reliability and stability of the fit, an additional constraint 
was applied to the equivalent width of the absorption component, by allowing 
it to vary only between 0 and 5 \AA.

\subsection{Maps construction}
\label{SubSect:MapsConstruction}

The emission line fit procedure gives for each line parameter (for instance
flux) a table with the fiber ID number, the measured value and the
acceptance/rejection flag. This table was then used to produce a 2D map, by
using an IRAF script that takes advantage of the fact that PMAS fibers are
arranged in a regular $16\times16$ matrix. The script automatically converts
ADU counts into flux (erg s$^{-1}$ cm$^{-2}$ A$^{-1}$), by multiplying by 
the wavelength dependent conversion factor taken from the sensitivity curve of 
the flux calibration.

Two continuum maps were obtained by summing the flux within specific 
wavelength intervals (4500 to 4700 \AA and 6000 to 6200 \AA), selected so as 
to avoid emission lines or strong residuals from the sky spectrum subtraction.
Line ratios maps were simply derived by dividing the corresponding flux maps.


\begin{table}
\caption[]{Basic data for Mrk~1418}
\label{Table:datamrk1418}
\begin{center}
\begin{tabular}{ll}     
\hline
\noalign{\smallskip}
Parameter      &  Value  \\
\noalign{\smallskip}
\hline
\noalign{\smallskip}
RA (2000)				       &  \phantom{+}09:40:27  	     \\
DEC (2000)			               &  +48:20:16		     \\
$v_\mathrm{helio}$ (km\,s$^{-1}$)	       &  $773\pm11$		     \\
$D$ (Mpc)				       &  $14.6\pm1.0$ 		     \\
$A_{B}$ (mag)$^\mathrm{a}$		       &  0.098		             \\
$m_{B}$ (mag)$^\mathrm{b}$		       &  $13.86\pm0.03$	     \\
$m_{R}$ (mag)$^\mathrm{b}$		       &  $12.61\pm0.19$	     \\
$M_{B}$ (mag)$^\mathrm{c}$		       &  $-16.96$		     \\
$F(\Ha)$ (ergs s$^{-1}$ cm $^{-2})^\mathrm{b}$ &  $95\pm16 \times 10^{-14}$  \\       
$m_{J}$ (mag)$^\mathrm{d}$		       &  $12.08\pm0.03$	     \\
$m_{H}$ (mag)$^\mathrm{d}$		       &  $11.46\pm0.04$	     \\
$m_{K_{s}}$ (mag)$^\mathrm{d}$  	       &  $11.41\pm0.05$	     \\
$M_\mathrm{HI}\;(M_{\sun})^\mathrm{e}$         &  $2.30\times10^{8}$         \\
\noalign{\smallskip}
\hline
\end{tabular}
\end{center}
Notes: RA, DEC, heliocentric velocity, $v_\mathrm{helio}$, and distance, $D$ 
(Mpc), all taken from NED (http://nedwww.ipac.caltech.edu/).
Distance calculated using a Hubble constant of 73 km s$^{-1}$ Mpc$^{-1}$, 
and taking into account the influence of the Virgo Cluster, the Great 
Attractor and the Shapley supercluster.
\begin{list}{}{}
\item[$^{\mathrm{a}}$] Absorption coefficient in the $B$ band, from 
\cite{Schlegel1998}.
\item[$^{\mathrm{b}}$] $B$- and $R$-band integrated magnitudes, their 
$1\sigma$ uncertainty and integrated \Ha\ flux,  
corrected for Galactic extinction \citep{GildePaz2003}.
\item[$^{\mathrm{c}}$] Absolute magnitude in the $B$ band, computed from 
the above tabulated $B$ integrated magnitude and distances.
\item[$^{\mathrm{d}}$] Total magnitudes from 2MASS.  
\item[$^{\mathrm{e}}$] Neutral hydrogen mass $M_\mathrm{HI}$ from 
\cite{VanZee2001}, which adopted a distance of 10.8 Mpc.
\end{list}
\end{table}

\section{Results}
\label{Sect:Results}

\subsection{Morphology of the ionized gas and stars}
\label{SubSect:Morphology}

\begin{figure*}
\centering
\includegraphics[width=10cm]{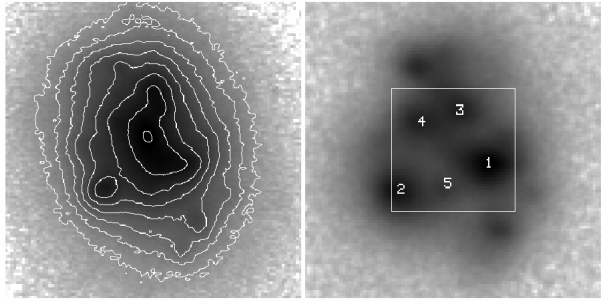}
\caption{Left panel: $B$-band image of Mrk~1418 with isocontours, spaced 0.5 
mag apart, overlaid. Right panel: continuum subtracted \Ha\ frame, with 
the $16 \times 16$ arcsec$^{2}$ PMAS field of view overlaid and the main 
star-forming knots labeled. North is up, east to the left. 
Both images are from \cite{GildePaz2003}, and publicly available in the NED;
the PSF is $2\farcs6$ FWHM.}
\label{Fig:images}
\end{figure*}

\begin{figure*}
\mbox{
\centerline{
\hspace*{0.0cm}\subfigure{\includegraphics[width=0.3\textwidth]{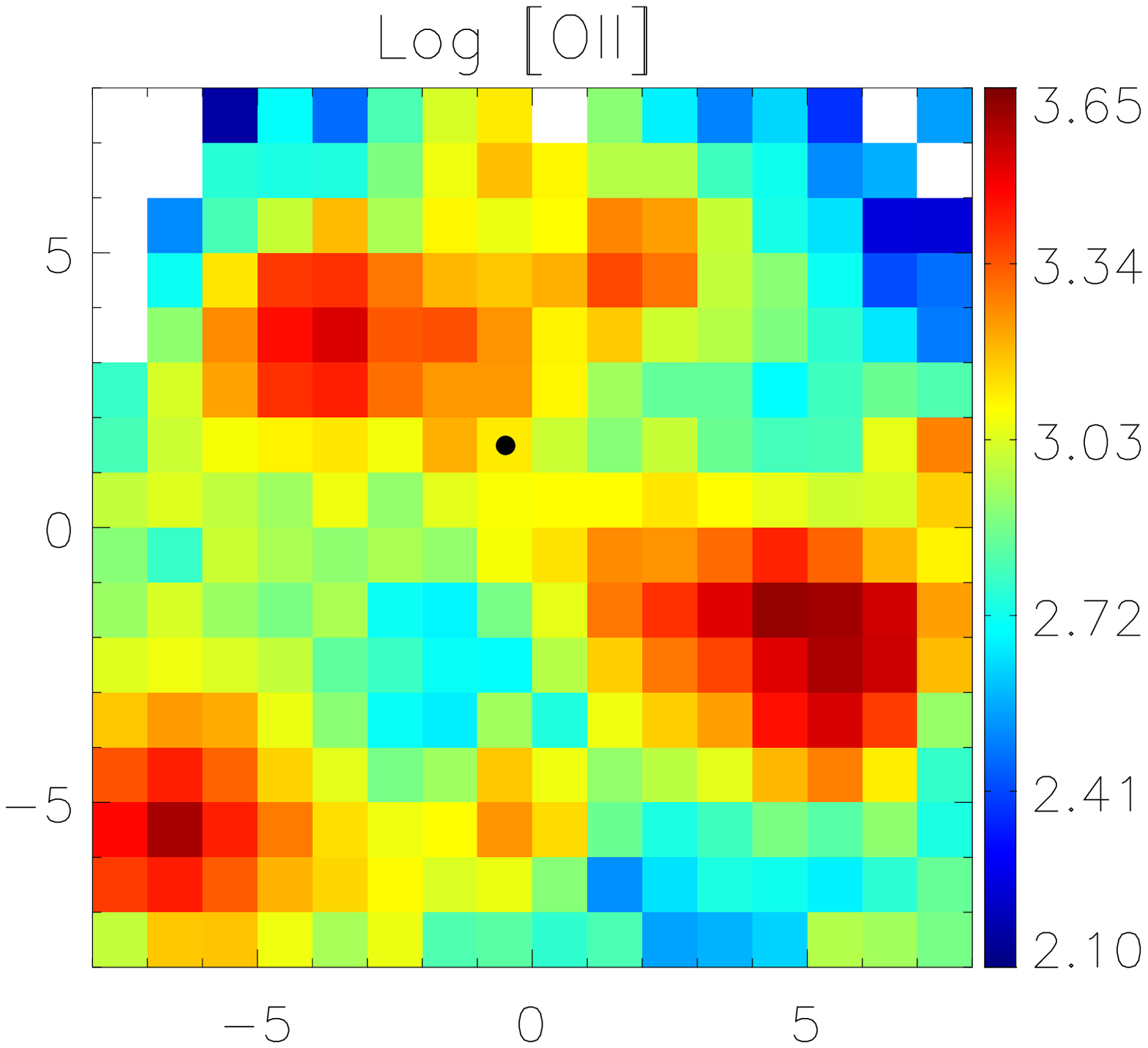}}
\hspace*{0.0cm}\subfigure{\includegraphics[width=0.3\textwidth]{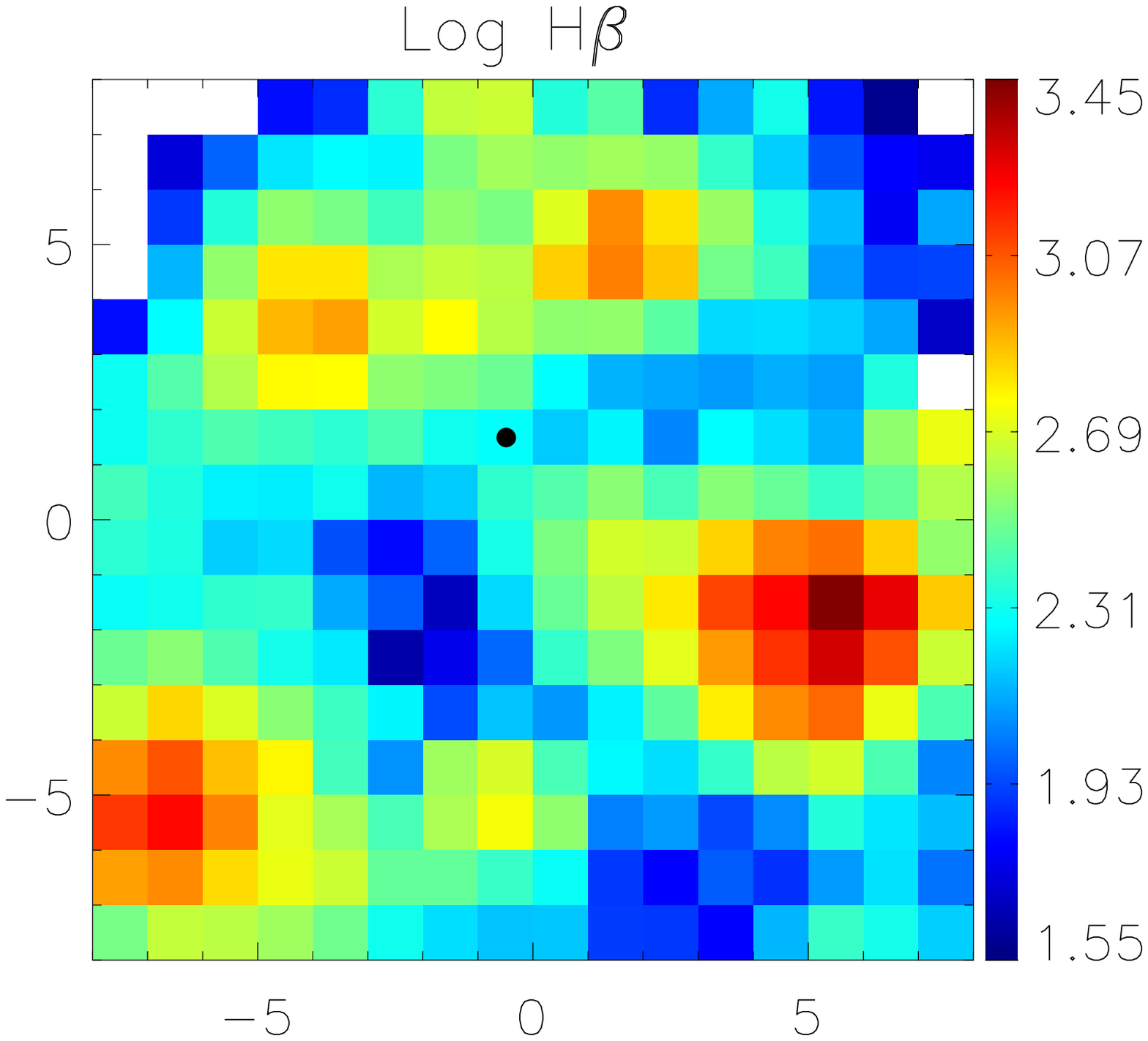}}
\hspace*{0.0cm}\subfigure{\includegraphics[width=0.3\textwidth]{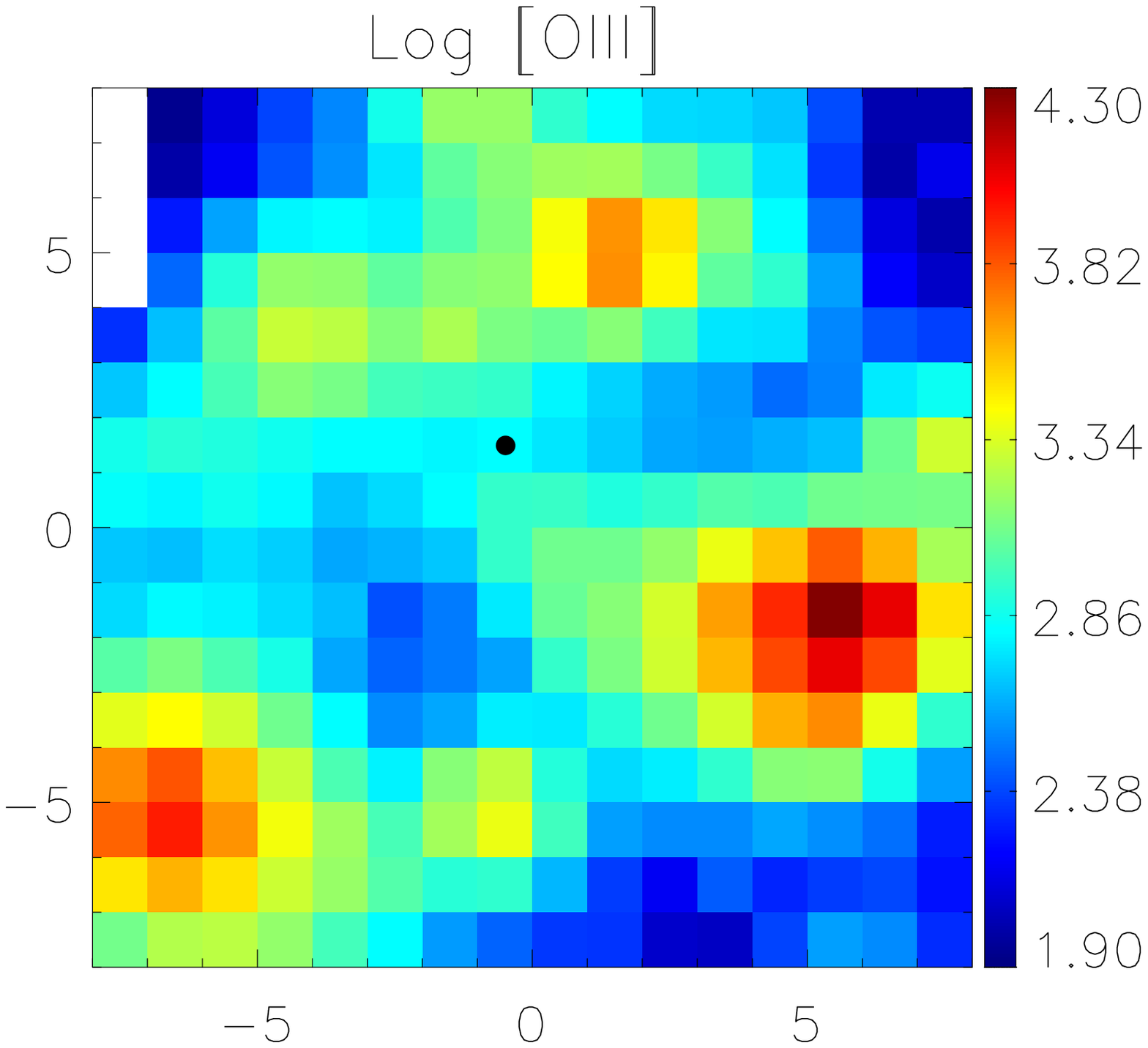}}
}}   
\mbox{
\centerline{
\hspace*{0.0cm}\subfigure{\includegraphics[width=0.3\textwidth]{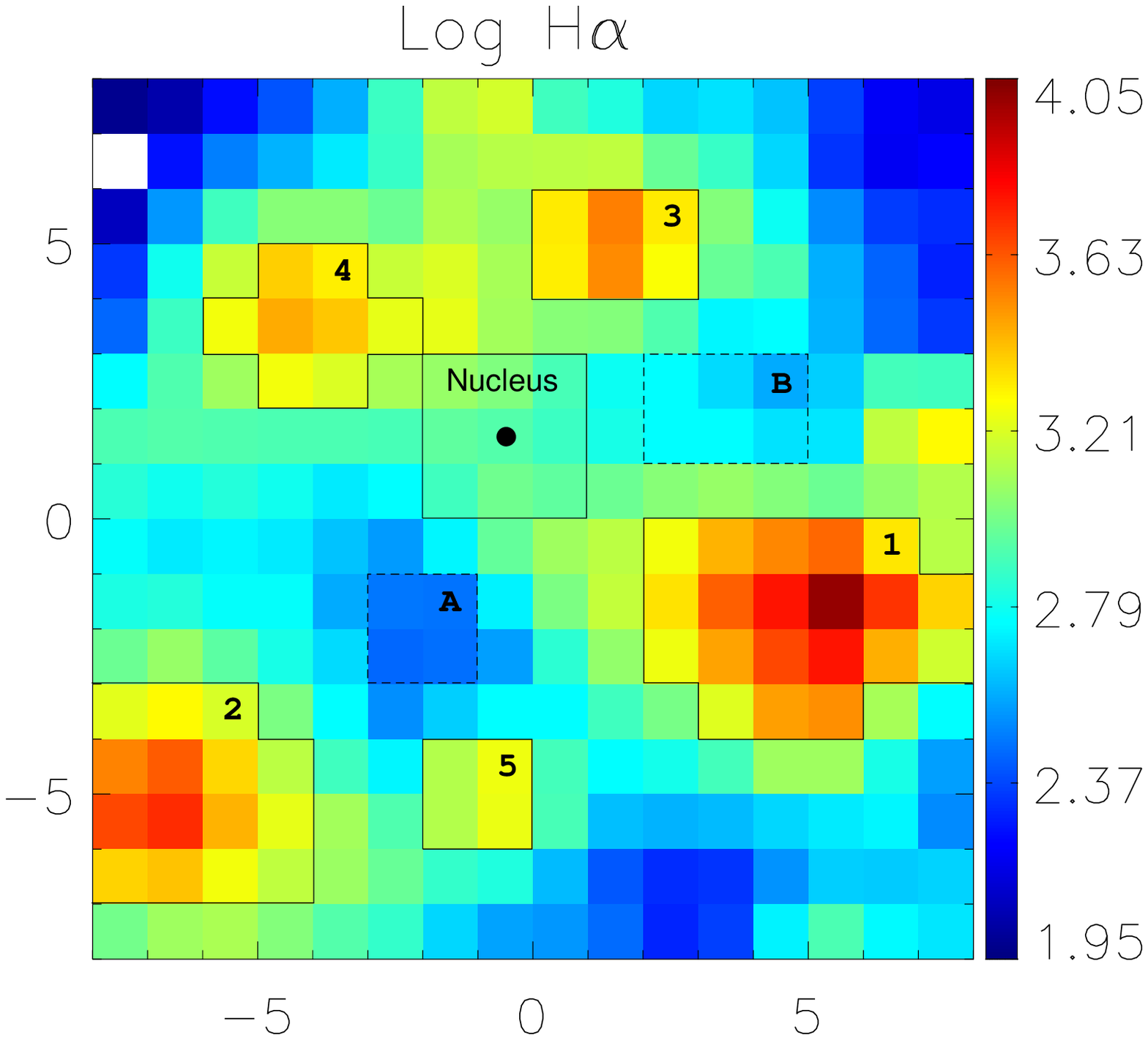}}
\hspace*{0.0cm}\subfigure{\includegraphics[width=0.3\textwidth]{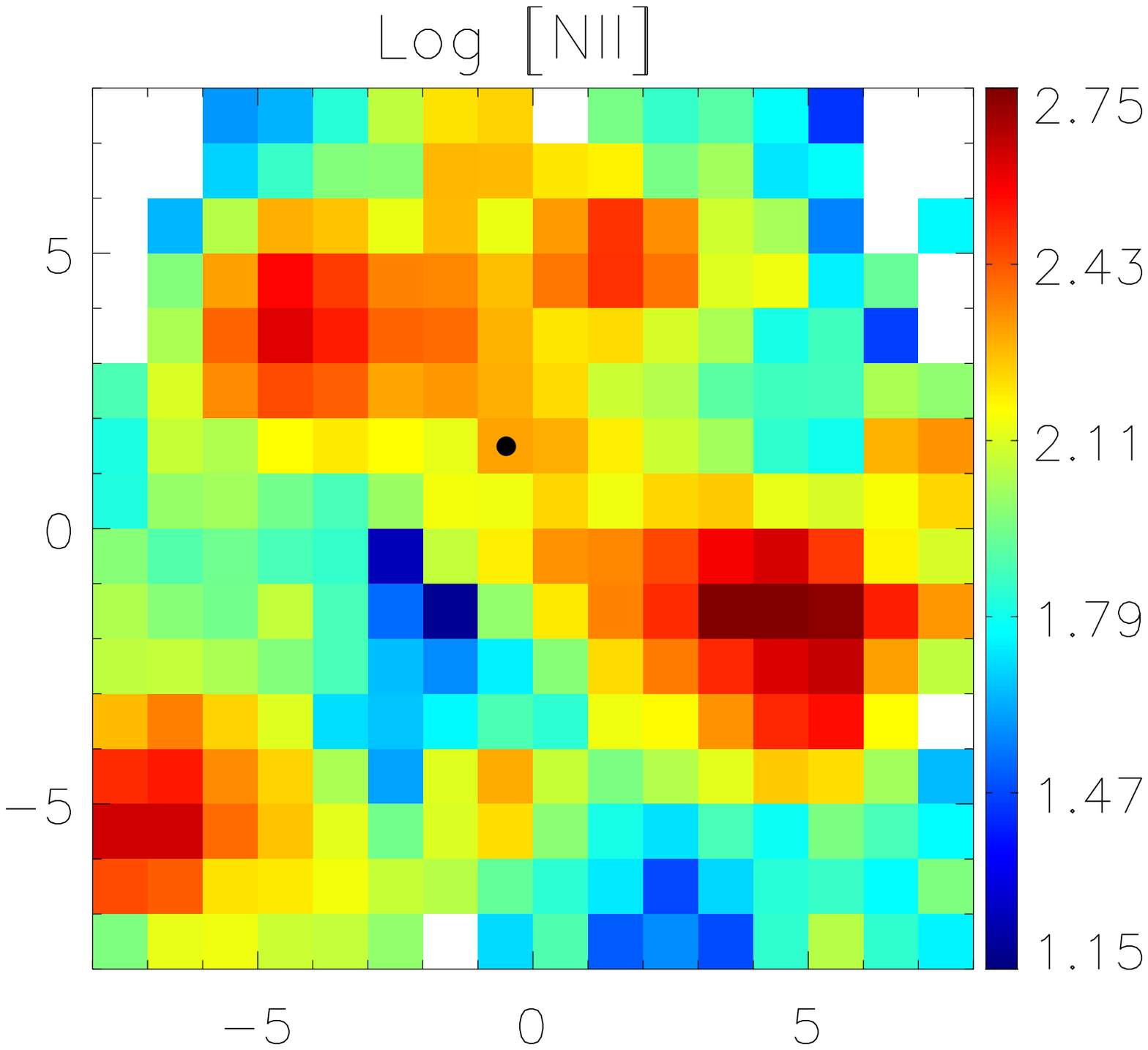}}
\hspace*{0.0cm}\subfigure{\includegraphics[width=0.3\textwidth]{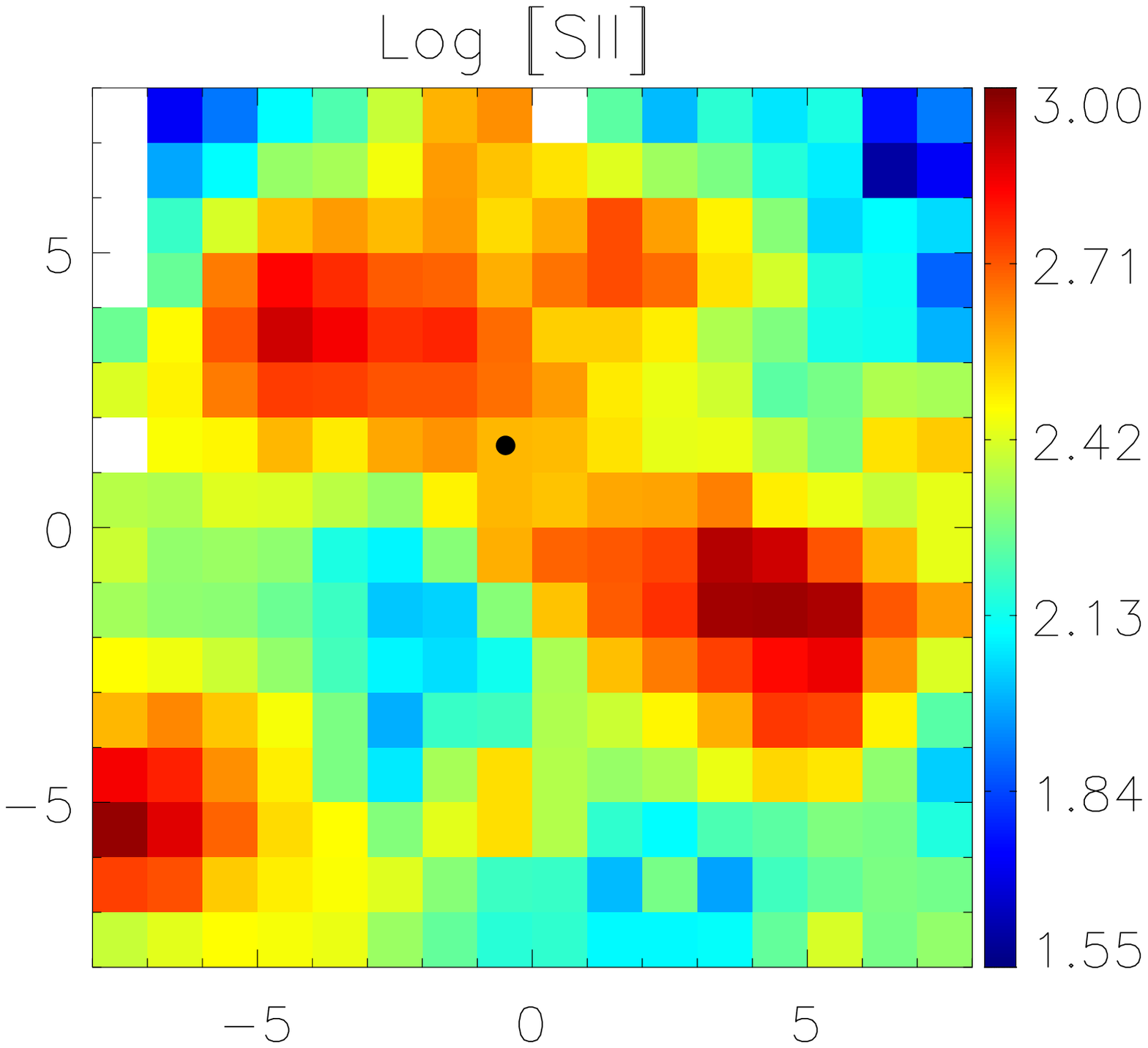}}
}} 
\mbox{ 
\centerline{
\hspace*{0.0cm}\subfigure{\includegraphics[width=0.3\textwidth]{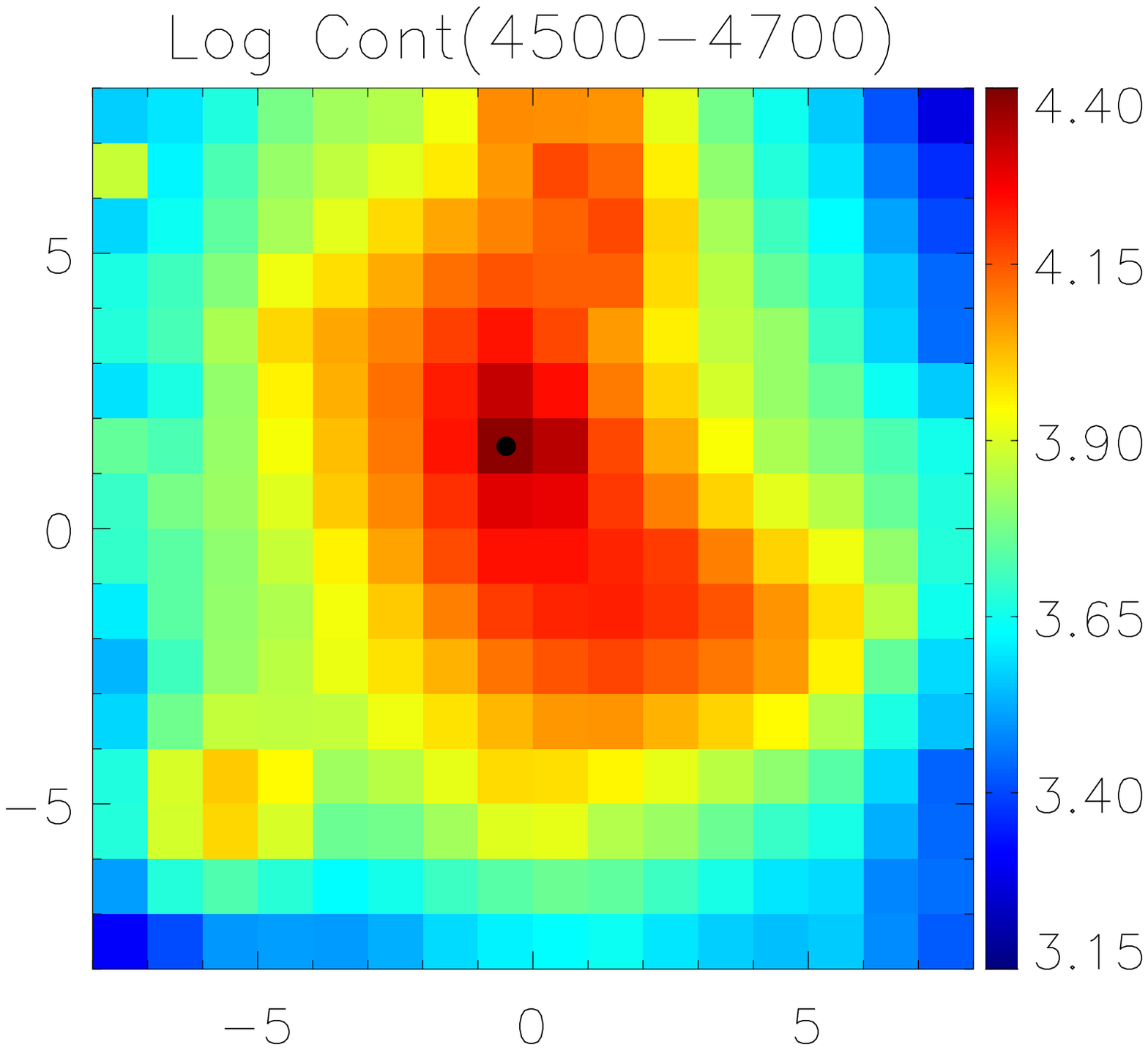}}
\hspace*{0.0cm}\subfigure{\includegraphics[width=0.3\textwidth]{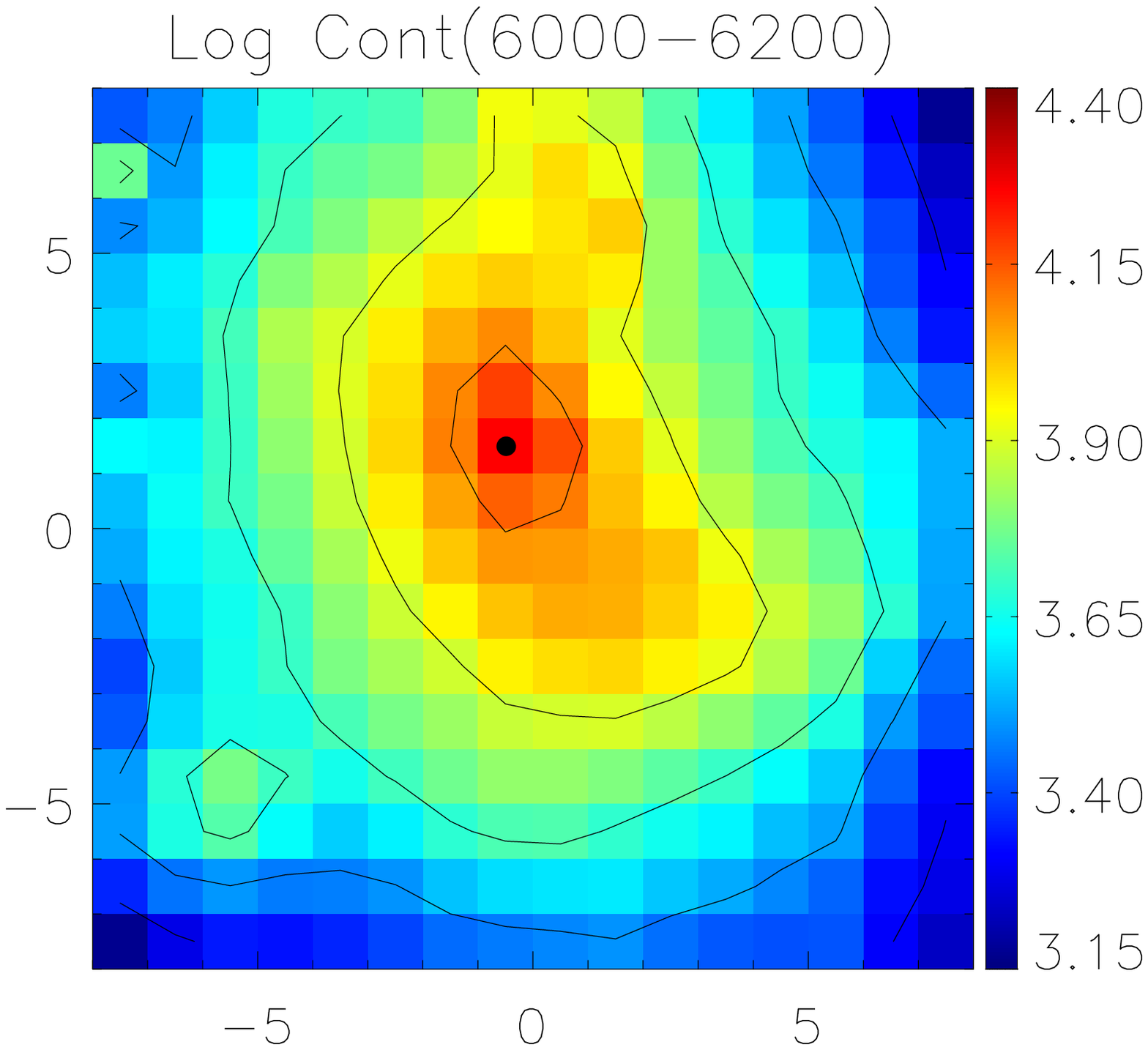}}
\hspace*{0.0cm}\subfigure{\includegraphics[width=0.3\textwidth]{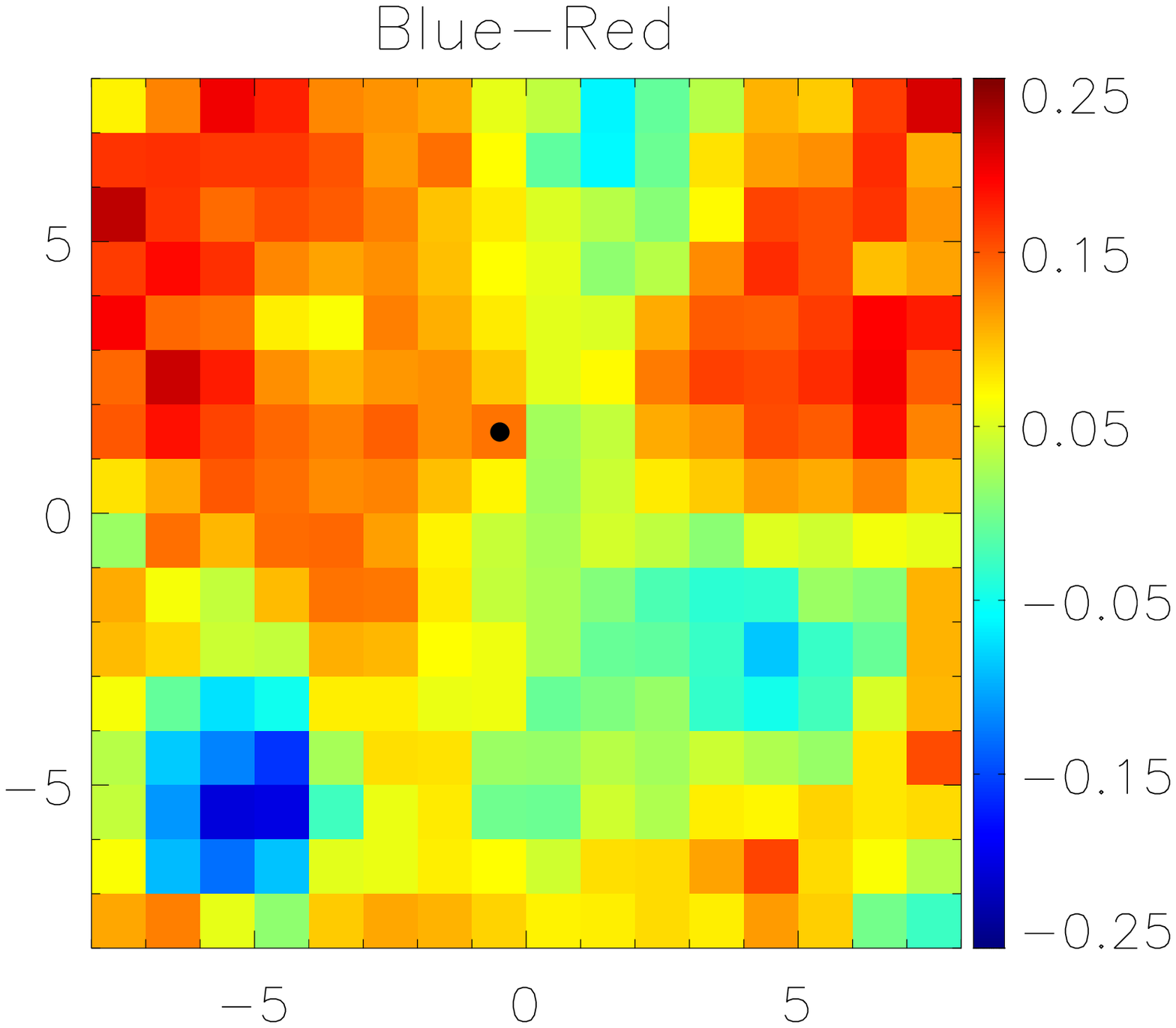}}
}} 
\caption{ Emission line maps for 
[\ion{O}{ii}]~$\lambda$3727, \Hb, [\ion{O}{iii}]~$\lambda5007$, \Ha, 
[\ion{N}{ii}]~$\lambda6584$ and [\ion{S}{ii}]~$\lambda\lambda6717,\;6731$,
continuum maps in the intervals 4500--4700 \AA\ (blue) and 6000--6200 \AA\
(red), and the continuum color map.
Axis units are arcseconds; north is up, east to the left. Maps are in  
logarithmic scale (in units of $10^{-18}$ erg s$^{-1}$ cm$^{-2}$), 
except the ``blue--red'' continuum color map which is in magnitudes (with 
arbitrary zero point). The black dot marks the nucleus of the galaxy, 
defined as the peak in the continuum frames.
The \Ha\ map also shows the outline of the SF knots and of the two 
"interknot" regions, A and B (see text for details).
The isophotes of the ``red'' continuum map are spaced 0.2 dex apart.}
\label{Fig:linefluxes}
\end{figure*}

To study the galaxy morphology we use flux maps of the brightest 
emission lines and the two continuum bands above.
The emission line maps trace the interstellar gas, ionized by massive stars, 
while the continuum maps show the stellar light distribution. 
For comparison, Figure~\ref{Fig:images} displays the published 
$B$ broad-band 
and \Ha\ images of Mrk~1418 \citep{GildePaz2003}, with the field of 
view covered by the PMAS observations overplotted. There is good qualitative 
agreement between the reconstructed maps (see Figure~\ref{Fig:linefluxes}) 
and the direct images.

Figure~\ref{Fig:linefluxes} displays the emission line flux maps in 
[\ion{O}{ii}]~$\lambda3727$, \Hb, [\ion{O}{iii}]~$\lambda5007$, \Ha,
[\ion{N}{ii}]~$\lambda6584$, and [\ion{S}{ii}]~$\lambda\lambda6717,\;6731$.
The galaxy shows the same morphology in all emission lines: five SF knots
are distributed in a roughly circular pattern around the optical center of the
galaxy. The emission peak is located in knot~1 in all the lines (see
Figure~\ref{Fig:images}), in the western part of the galaxy. This large knot
seems to be connected with knot~4, placed $\approx 10\arcsec$ 
to the northeast, with a bridge-like structure.

In the continuum (see bottom panels in Figure~\ref{Fig:linefluxes}), Mrk~1418
shows an inner kidney-shaped feature, with a peak located approximately at
the center of the outer isophotes. The position of this central maximum in the
continuum frames does not depend on what specific ``emission line free''
spectral range is used, hence we can safely define this continuum peak as the
optical nucleus of the galaxy. Smaller condensations  are detected north and
southeast of the galaxy nucleus. 

The structure of the ionized gas, as traced by the emission lines, looks 
quite different from the stellar light distribution. At the position of the 
continuum peak, there is a dearth of gas emission, indicating that the star 
formation in the central regions has already ceased. The position of the 
knots detected in emission does not coincide with any continuum peak
(knot~2 is close, but not coincident with the continuum blob seen in the 
southeast corner).
Knots~3 and 4 seem to be co-located with a shallow, local blue maximum 
in the "blue--red" continuum map, while knots~1 and 2 are slightly displaced 
from the blue continuum peaks.

\subsection{Ionization: mechanism and structure}
\label{SubSect:Ionization}

\begin{figure*}
\mbox{
\centerline{
\hspace*{0.0cm}\subfigure{\includegraphics[width=0.3\textwidth]{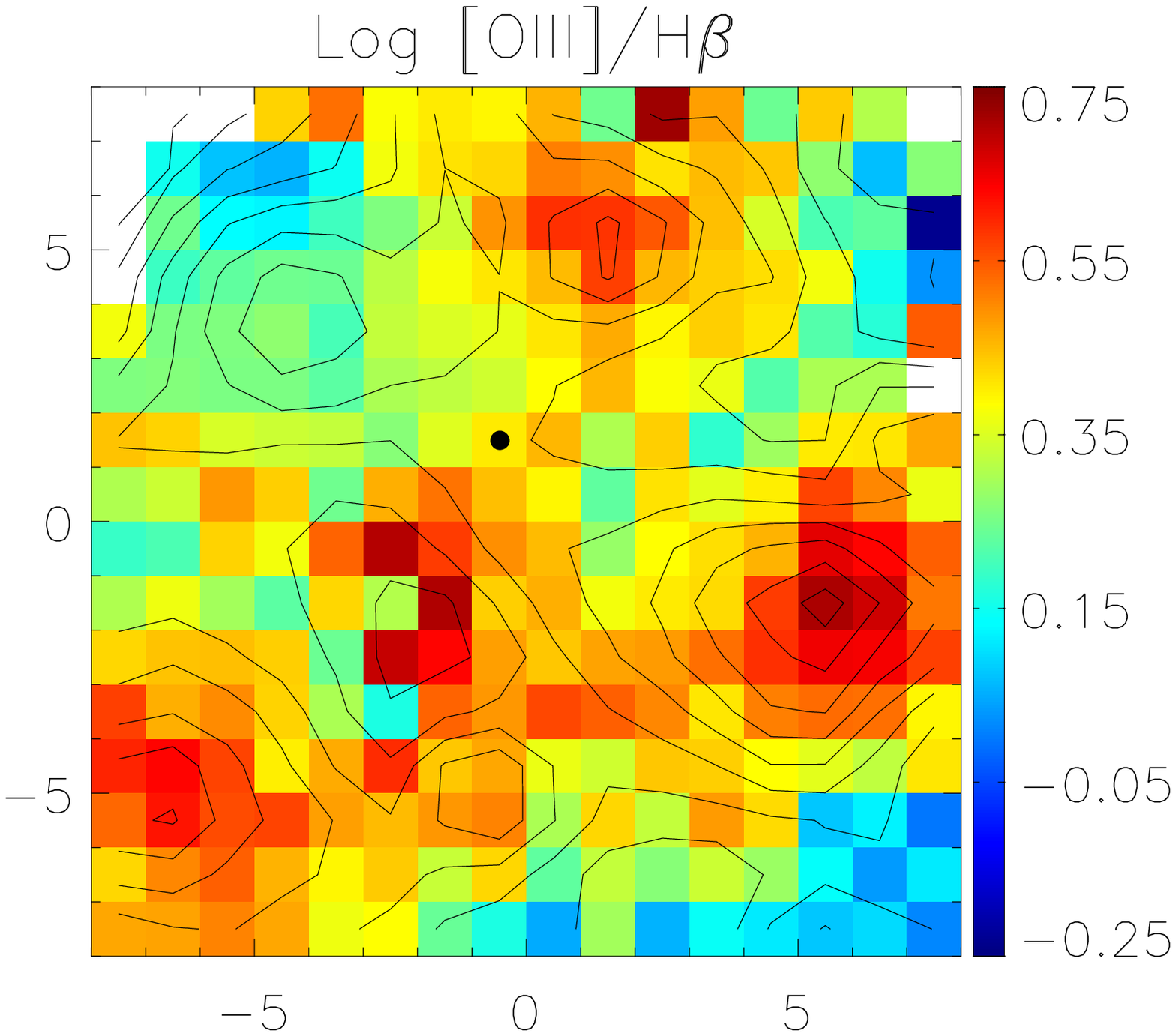}}
\hspace*{0.0cm}\subfigure{\includegraphics[width=0.3\textwidth]{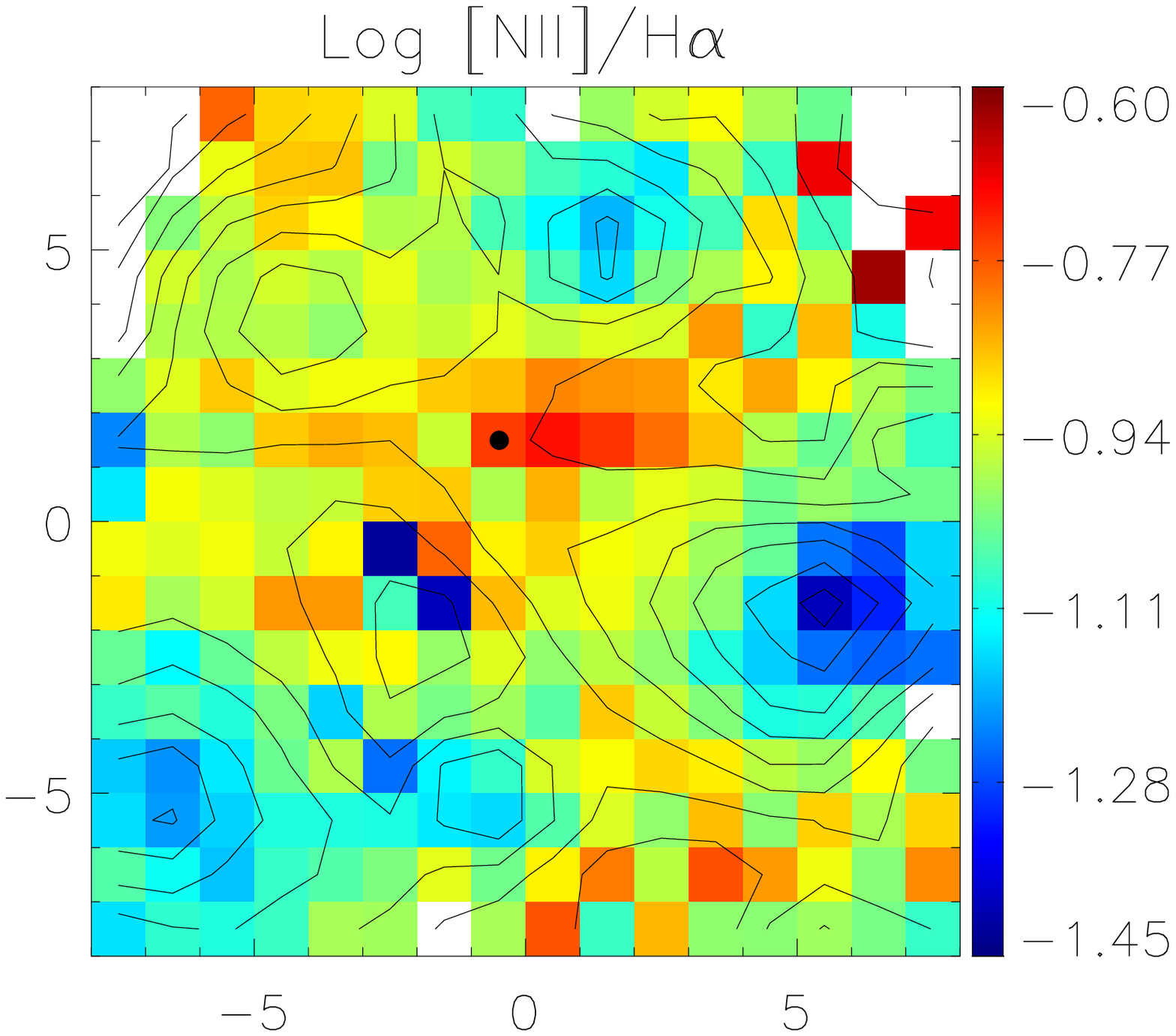}}
\hspace*{0.0cm}\subfigure{\includegraphics[width=0.3\textwidth]{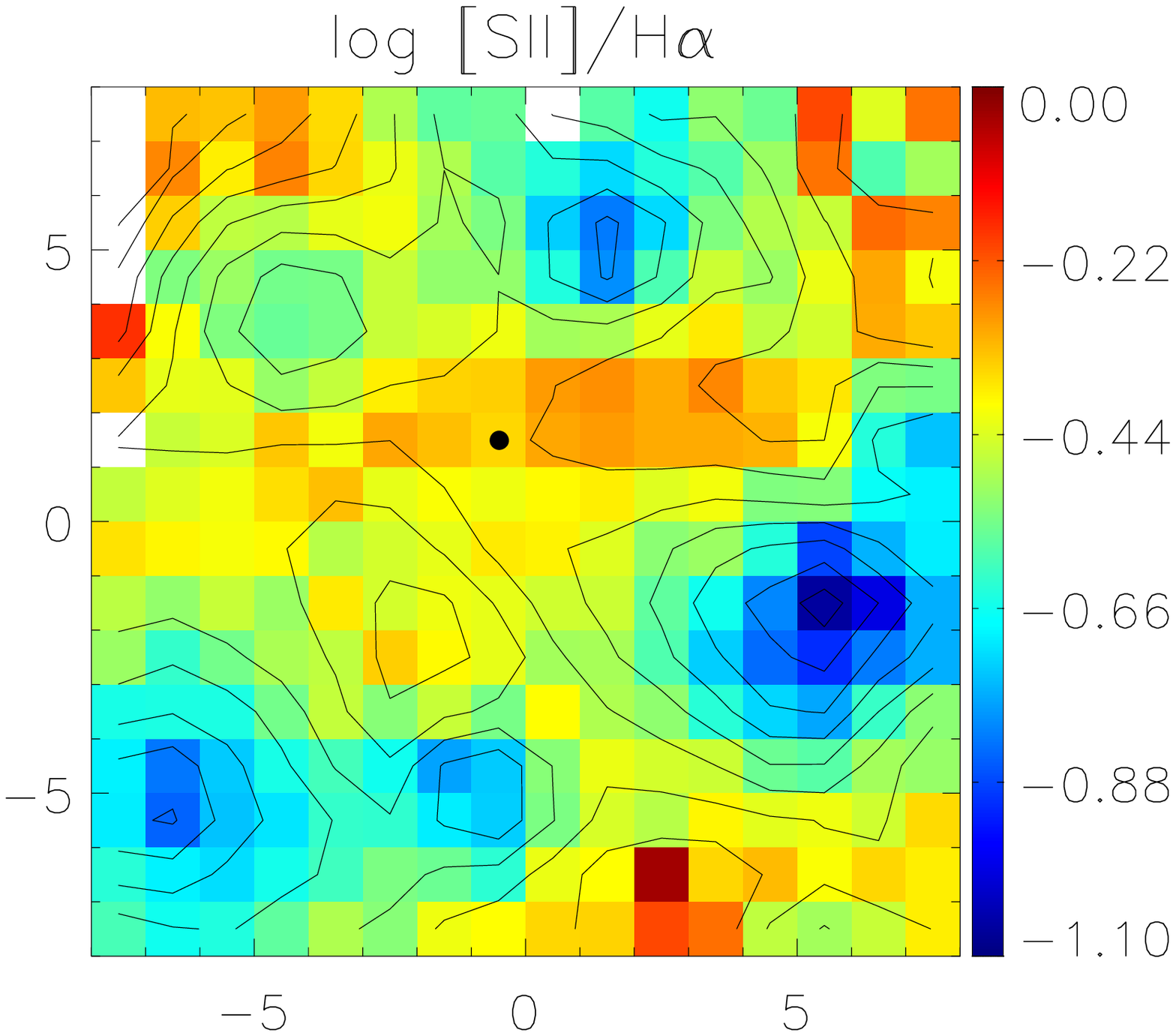}}
}}   
\mbox{
\centerline{
\hspace*{0.0cm}\subfigure{\includegraphics[width=0.3\textwidth]{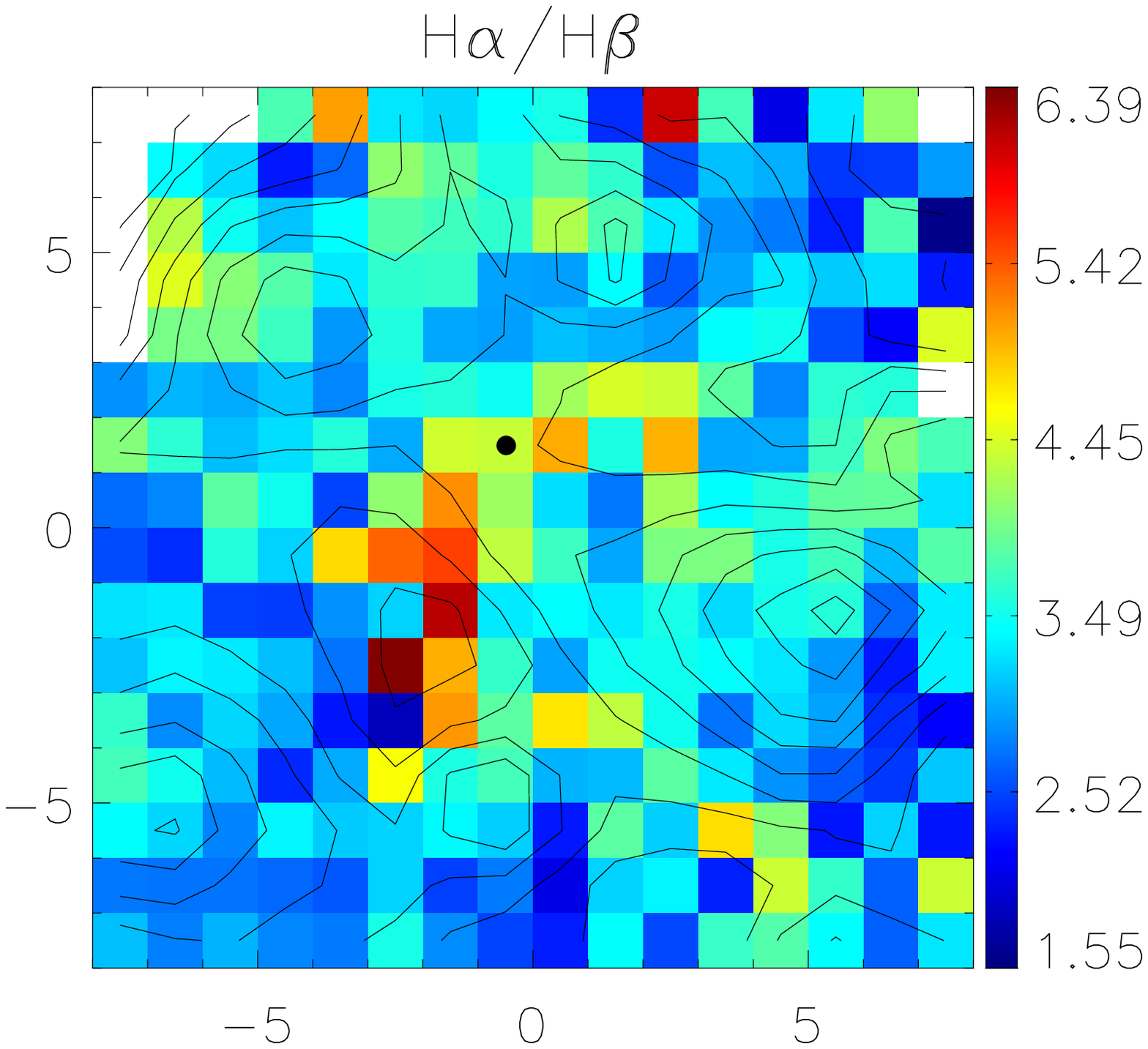}}
\hspace*{0.0cm}\subfigure{\includegraphics[width=0.3\textwidth]{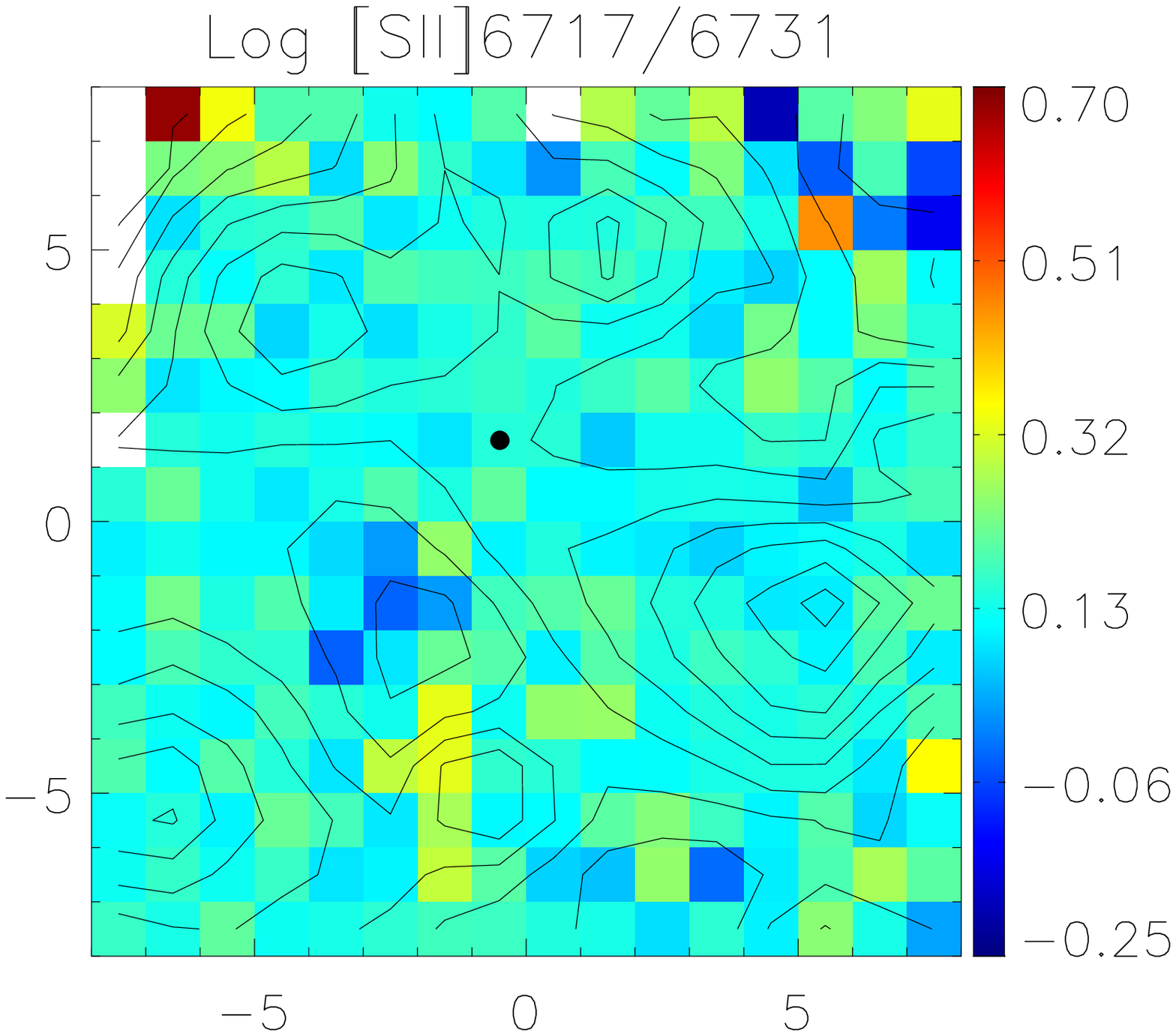}}
\hspace*{0.0cm}\subfigure{\includegraphics[width=0.3\textwidth]{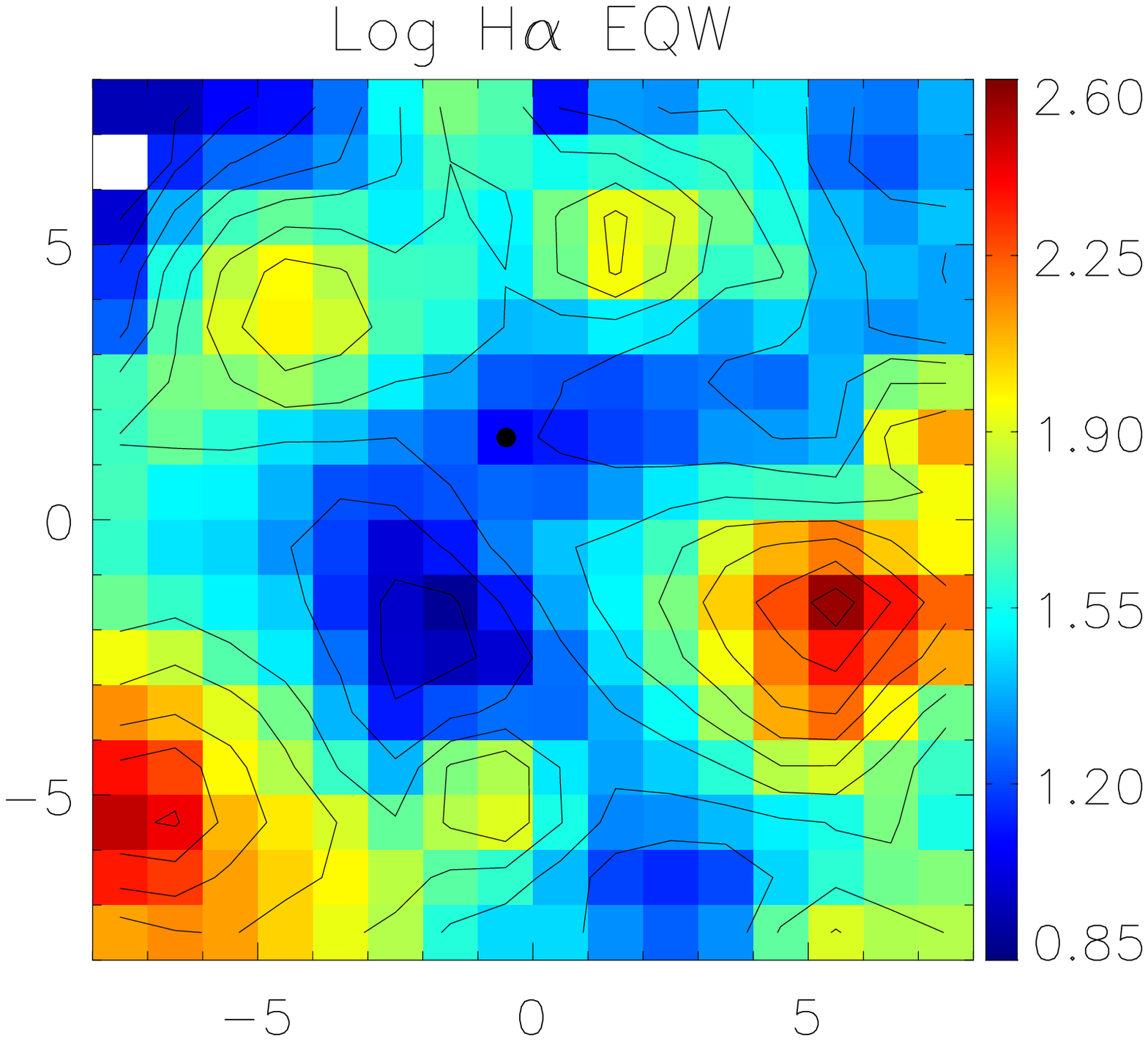}}
}}
\caption{Line ratios maps: [\ion{O}{iii}]~$\lambda5007$/\,\Hb,  
[\ion{N}{ii}]~$\lambda6584$/\,\Ha,
[\ion{S}{ii}]~$\lambda\lambda6717,\;6731$/\,\Ha,     
\Ha/\,\Hb\, and
[\ion{S}{ii}]~$\lambda6717$/[\ion{S}{ii}]~$\lambda6731$.      
The last panel shows the \Ha\ equivalent width map. 
\Ha\ contours (spaced 0.2 dex apart) have been overlaid in all the maps. 
Axis units are arcseconds; north is up, east to the left. 
All maps are in logarithmic scale, except \Ha/\Hb.
The black dot marks the nucleus of the galaxy.}
\label{Fig:lineratios}
\end{figure*}

To obtain information about the ionization structure of the ionized nebulae 
and the nature of the ionization sources acting in Mrk~1418, we built 
emission line ratio maps in [\ion{O}{iii}]~$\lambda5007$/\Hb,
[\ion{N}{ii}]~$\lambda6584$/\Ha\, and
[\ion{S}{ii}]~$\lambda\lambda6717,\;6731$/\Ha.  Although
[\ion{O}{i}]~$\lambda6300$ is detectable in many spaxels, it unfortunately 
falls very close to the sky line [\ion{O}{i}]~$\lambda6300$, which prevents 
us from obtaining precise measurement of its flux and, hence, of the
[\ion{O}{i}]~$\lambda6300$/\Ha\ ratio.

Using the information derived from the above mentioned line ratios it is
possible to distinguish among the different ionization mechanisms acting
in a gaseous nebula, namely, photoionization by radiation from OB stars,
shock-wave heating and photoionization by a power law continuum source
\citep{Baldwin1981,Veilleux1987}. 
The [\ion{O}{iii}]~$\lambda5007$/\Hb\ emission line ratio is an
excitation parameter indicator and provides information about the available 
fraction of hard ionizing photons of the ionizing star cluster embedded in the 
nebula;  therefore a large [\ion{O}{iii}]~$\lambda5007$/\Hb\
ratio indicates a highly ionized region. [\ion{N}{ii}]~$\lambda6584$ and
[\ion{S}{ii}]~$\lambda\lambda6717,\;6731$ are low ionization lines, usually 
weak in \ion{H}{ii} regions: large [\ion{N}{ii}]~$\lambda6584$/\Ha\ and
[\ion{S}{ii}]~$\lambda\lambda6717,\;6731$/\Ha\ ratios usually indicate the
presence of ionizing mechanisms different from photoionization, e.g. shock 
waves and/or AGNs (e.g. \citealp{Shull1979}).

Figure~\ref{Fig:lineratios} displays the maps of 
[\ion{O}{iii}]~$\lambda5007$/\Hb, [\ion{N}{ii}]~$\lambda6584$/\Ha\ and
[\ion{S}{ii}]~$\lambda\lambda6717,\;6731$/\Ha. In order to allow a quick
comparison with the position of the ionization sources, \Ha\ contours have
been overlaid. These three maps show basically the
same structure. As expected in regions photoionized by stars, they trace the
star-formation activity, and the peaks in [\ion{O}{iii}]~$\lambda5007$/\Hb\
coincide with the minima in [\ion{N}{ii}]~$\lambda6584$/\Ha\ and
[\ion{S}{ii}]~$\lambda\lambda6717,\; 6731$/\Ha. Only knot~4 displays a
different behavior: it is a moderate minimum in 
[\ion{N}{ii}]~$\lambda6584$/\Ha\ and
[\ion{S}{ii}]~$\lambda\lambda6717,\;6731$/\Ha\, and also shows low values in
[\ion{O}{iii}]~$\lambda5007$/\Hb. The [\ion{O}{iii}]~$\lambda5007$/\Hb\ map
also displays a peak at about $(-2\arcsec,-2\arcsec)$, the same 
position as  the peak in the extinction map, in an area that does not 
coincide with any emission knot.

\subsection{Extinction}
\label{SubSetc:Extinction}

The extinction map was built from the \Ha/\Hb\ ratio. 
No reliable fits could be obtained, for the individual spaxels, for 
\Hd\ and \Hg, as they are much weaker and are also strongly affected by 
underlying stellar absorption.

The \Ha/\Hb\ ratio map is shown in Figure~\ref{Fig:lineratios}. Mrk~1418 shows
an inhomogeneous extinction pattern, which peaks in the central, inter-knot
regions, displaced from the continuum maximum ($\approx 3\arcsec$ to the 
southeast). There, \Ha/\Hb\ varies between 4 and 5.5, suggesting
strong extinction by dust with a reddening $E(B-V)$ of up to 0.62 mag, while
in the surrounding SF knots the \Ha/\Hb\ ratio is  $\leq3$, that is
$E(B-V)\leq 0.04$ mag.
Indeed, an extinction patch or lane located between the emission line 
knots has been already observed in several BCDs (e.g. \citealt{GarciaLorenzo2008,
Vanzi2008,Kehrig2008}), and could be related to the sweeping out of the ISM 
by stellar winds and SNs. 
In the continuum "blue--red" color map, the position of the extinction peak 
is redder than the knots, but no dust lane or patch is clearly seen;
changes in stellar populations as well as the specific geometry of the dust 
distribution may account for this seeming incongruity.

This result stresses the importance of performing a bidimensional correction
of the interstellar extinction: just assuming a single, spatially constant
value for the extinction can lead to large errors in the derivation of fluxes
and magnitudes in the different regions of the galaxy. 
As an example, if we used the extinction value for knot~1, $E(B-V)\sim
0.07$, to correct the observed colors in the nuclear region (where the
extinction is $E(B-V) \sim 0.40$) we would obtain a color $B-V$ about 0.3 mag
redder than its actual value.

\subsection{Electron density distribution}
\label{SubSect:ElectronDensity}

We also produced the map of the ratio
[\ion{S}{ii}]~$\lambda6717$/[\ion{S}{ii}]~$\lambda6731$, an electron
number density sensitive diagnostic for the range 100--10.000 cm$^{-3}$
(Figure~\ref{Fig:lineratios}). We find that the density is almost constant,
with values close to the low density limit in almost the whole field of view.
Only in a few fibers in the central, inter-knot region (the same
one where the peaks of the [\ion{O}{iii}]~$\lambda5007$/\Hb\ ratio and of
the extinction are located) the density is higher.

\subsection{Integrated spectroscopy of selected galaxy regions}
\label{SubSect:IntegratedSpectroscopy}

As the next step in our spectroscopic analysis, we used the derived flux
and continuum maps to isolate the regions of interest in the central parts of
the galaxy. They include the five brightest SF knots detected in the
narrow-band maps and the nucleus, defined as the continuum peak (see
Figure~\ref{Fig:linefluxes}).

We identified the fibers corresponding to each region, and summed their
spectra together to create the final spectrum. The regions were 
defined as those spaxels with an \Ha\ flux larger than 3.2 (in logarithm),
a threshold chosen as the best compromise between enclosing a sufficient
number of spaxels for each knot, and avoiding that the knots overlap or abut
each other. A few spaxels above this threshold but whose inclusion would have 
stretched the shape of a knot were excluded, while knot~5 was augmented by
two additional spaxels below the threshold. 

We also defined two interknot regions: A, where the \Ha\ flux has a
local minimum, and B, where the
[\ion{S}{ii}]~$\lambda\lambda6717,\;6731$/\Ha\ ratio is highest (see
Figure~\ref{Fig:linefluxes}).

In this way we produced higher signal-to-noise spectra (in comparison to the
spectra of the individual fibers), better suited to derive physical parameters
and chemical abundances. The area of these regions ranges between $\sim 0.02$
and 0.10 kpc$^{2}$. We also computed the total spectrum of Mrk~1418 by summing
all the 256 spaxel spectra (that is, over an area of 1.3 kpc$^{2}$).

\begin{figure}
\centering
\includegraphics[angle=0, width=0.5\textwidth]{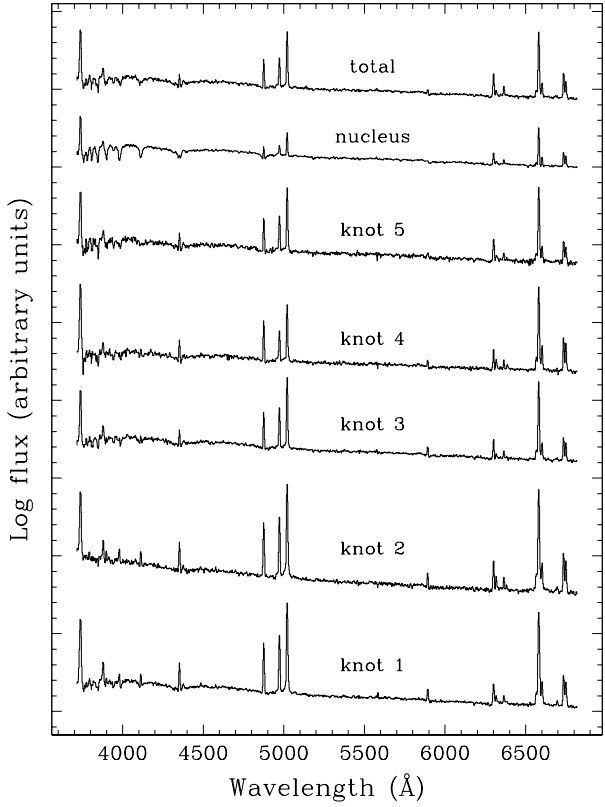}
\caption{Spectra of the 5 knots, of the nucleus, and the total spectrum 
obtained by summing all the fibers. Spectra are shown in logarithmic scale 
and are offset for clarity. Small tickmarks are spaced 0.2 dex apart.}
\label{Fig:spectra}
\end{figure}

Fig.~\ref{Fig:spectra} displays the spectra for the six selected regions and
the whole mapped field of view. The spectra of the five SF regions show
similar characteristics: all of them are dominated by strong emission
features, superimposed on a  blue continuum. Strong
[\ion{O}{ii}]~$\lambda3727$, \Hb, [\ion{O}{iii}]~$\lambda5007$, \Ha,
[\ion{N}{ii}]~$\lambda6584$ and [\ion{S}{ii}]~$\lambda\lambda6717,\;6731$, are
visible in emission. Absorption wings are detected in \Hd, \Hg\ and \Hb, for
knots 3, 4, and 5. Although [\ion{O}{i}]~$\lambda6300$ is detected in the
spectra of all the five knots, no reliable flux measurements could be
obtained, as the profile of this line  is contaminated by the residuals from
the subtraction of the [\ion{O}{i}]~$\lambda6300$ sky-line. The spectrum of
the nucleus shows strong absorption features, indicating an important
contribution of older stars; \Hd, \Hg\ and higher order Balmer lines (H11,
H10, H9, H8, H7) are visible only in absorption.

\subsubsection{Reddening corrected line fluxes} 
\label{SubSubSect:LineFluxes}

Fluxes and equivalent widths of the emission lines were measured by fitting
the line profile with a Gaussian. Reddening-corrected  intensity ratios and
equivalent  widths  for the different spatial regions are listed  in
Table~\ref{Table:fluxes_a}. The correction from underlying stellar absorption
was done in two different ways, as explained below.

In knots 4 and 5, and in the nuclear and the integrated spectra, the 
\Hb\ Balmer line shows clear absorption wings, and we fitted simultaneously
an absorption and an emission component.

In the remaining cases, we proceeded as follows. We first adopted an initial
estimate for the absorption equivalent width, EW$_\mathrm{abs}$, corrected the
measured fluxes, and computed the extinction coefficient $C(\Hb)$ through a 
least-square fit to the Balmer decrement given by the equation:

\begin{equation}
\label{Eqn:redde}
\frac{F(\lambda)}{F(\Hb)}=\frac{I(\lambda)}{I(\Hb)} 
\times 10^{C({\rm H}\beta)\times f(\lambda)} 
\end{equation}

\noindent where $\frac{F(\lambda)}{F({\rm H}\beta)}$ is the line flux 
corrected for absorption and normalized to \Hb; $\frac{I(\lambda)}{I({\rm
H}\beta)}$ is the theoretical value for case B recombination, from
\citet{Brocklehurst1971}, and $f(\lambda)$ is the reddening curve normalized
to \Hb\ which we took from \citet{Whitford1958}.

We then varied the  value of  EW$_\mathrm{abs}$, until we find the one that
provides the best match (e.g. the minimum scatter in the above relation)
between the corrected and the theoretical line ratios. This is done
separately for each knot.

We note that this correction accounts for the fraction of emission line flux
that is masked by the underlying absorption, and not necessarily represents
the actual absorption equivalent width. To illustrate this point, let us
consider two (extreme) examples: (1) the width of  the emission line and
that of  the absorption line  are very  similar; in this  case the absorption
component is completely  hidden underneath  the  stronger  emission 
component,  and  the measured emission flux will be the algebraic sum  of
the fluxes of  the two components (the flux of the absorption line being
negative); (2) the absorption line is very broad and shallow; in this case,
the emission line flux measured by setting the continuum level on both
sides of the ``base'' of  the emission line profile will only slightly
underestimate the actual flux, and the correction will be much smaller that
the actual equivalent width of the absorption component.

\begin{figure}
\centering
\includegraphics[angle=0, width=0.48\textwidth]{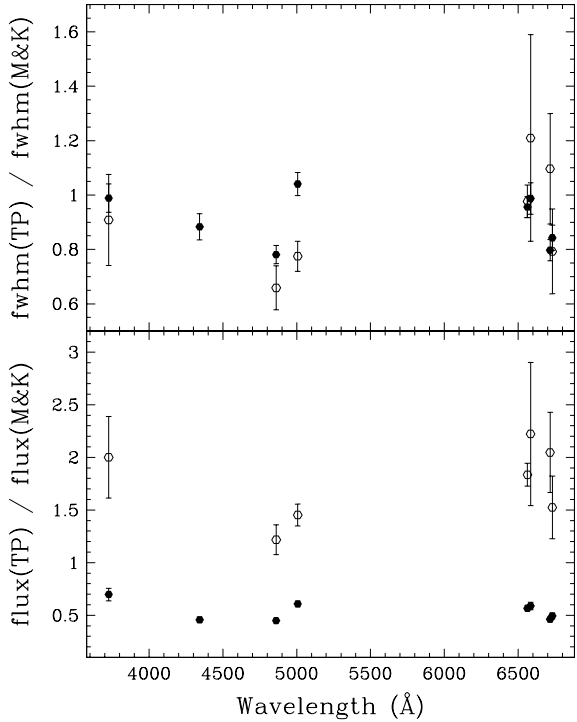}
\caption{Comparison between the integrated and nuclear fluxes and 
equivalent widths from this paper (TP), and the values published by 
\cite{MoustakasKennicutt2006} (M\&K), which used apertures of
$30\arcsec\times50\arcsec$ and $2\farcs5 \times 2\farcs5$ respectively.
For this comparison, our measurements have only been corrected for galactic 
extinction and, for the Balmer lines, for underlying absorption (see text for 
details). Filled circles represent the integrated spectra, open circles
correspond to the nuclear spectra. The line at $\simeq 4350$ \AA\ is \Hg.}
\label{Fig:Moustakas}
\end{figure}

We compared our measurements for the integrated and the nuclear
spectra with the spatially integrated (within a $30\arcsec \times 50\arcsec$
area) and nuclear ($2\farcs5 \times 2\farcs5$) values published by
\cite{MoustakasKennicutt2006}. The results are presented in
Fig.~\ref{Fig:Moustakas}: keeping into account the different sizes of the
apertures used, and a likely small spatial offset in the location of the
nuclear aperture, the comparison looks fair.

\subsubsection{Line ratios, physical parameters and chemical abundances}
\label{SubSubSect:LineRatios}

Table~\ref{Table:Ratios} lists the most relevant line ratios, physical
parameters and chemical abundances for the selected regions in Mrk~1418. In
order to calculate the errors associated with the physical properties and
chemical abundances, we propagated the emission-line intensity errors listed
in Table~\ref{Table:fluxes_a}.

The physical conditions (electron density and temperature) and ionic element
abundances were derived  from the  reddening-corrected emission line fluxes
following the 5-level atom {\em fivel} program in the IRAF  
\textsc{nebular} package \citep{deRobertis1987,ShawDufour1995}.

Electron densities were  calculated from the ratio of
[\ion{S}{ii}]~$\lambda$6717/$\lambda$6731  emission lines. The electron
temperature of [\ion{O}{iii}] ($T_\mathrm{e}$[\ion{O}{iii}]), representative of 
the high-excitation  zone  of  the  ionized   gas, was derived  from the
[\ion{O}{iii}] $\lambda $4363/($\lambda$4959+$\lambda$5007) ratio for knots 1
and  2, and for the summed spectrum. In order to calculate
$T_\mathrm{e}$[\ion{O}{ii}] we used the relation between [\ion{O}{ii}]
and [\ion{O}{iii}] electron temperatures from \cite{Pilyugin2006}. We assumed 
the approximation  $T_\mathrm{e}$[\ion{S}{ii}] $\approx  
T_\mathrm{e}$[\ion{N}{ii}] $\approx T_\mathrm{e}$[\ion{O}{ii}] for  
the calculation of S$^{+}$  and N$^{+}$ abundances,
since no auroral line could be  measured in the low excitation zone. We
adopted $T_\mathrm{e}$[\ion{O}{iii}] for the calculation  of  O$^{2+}$. As the
nebular \ion{He}{ii}~$\lambda4686$ \AA\  line was not detected in  any of the 
knot spectra, we can assume that the contribution from highly ionized  
species like O$^{3+}$ is negligible. Therefore, the total oxygen and 
nitrogen abundances were obtained as:
$\mathrm{O/H} = (\mathrm{O}^{+} + \mathrm{O}^{2+})/\mathrm{H}^{+}$
and
$\mathrm{N/O} = \mathrm{N}^{+}/\mathrm{O}^{+}$.

To obtain oxygen abundances in those knots in which [\ion{O}{iii}] could not
be  measured (knots~3, 4, 5, and the nucleus), we applied the
commonly  used  strong-line  method from \cite{PettinPagel2004}.  These
authors  revised the 
$N2$ ($\equiv$ log{[\ion{N}{ii}]~$\lambda6584$/\Ha}) and  
$O3N2$ ($\equiv$
log\{([\ion{O}{iii}]$\lambda5007$/\Hb)/([\ion{N}{ii}]$\lambda6584$/\Ha)\})
indices, using 137 extragalactic \ion{H}{ii} regions. The uncertainties in the 
derived metallicities from these two indices are $\sim 0.38$ and 0.25 
dex respectively.

For knots~1 and 2, the estimated oxygen abundances from empirical parameters
are consistent, within the uncertainties, with those derived using the
electron temperature.

\begin{figure}
\includegraphics[bb=12 83 563 448,angle=-90,width=0.28\textwidth]{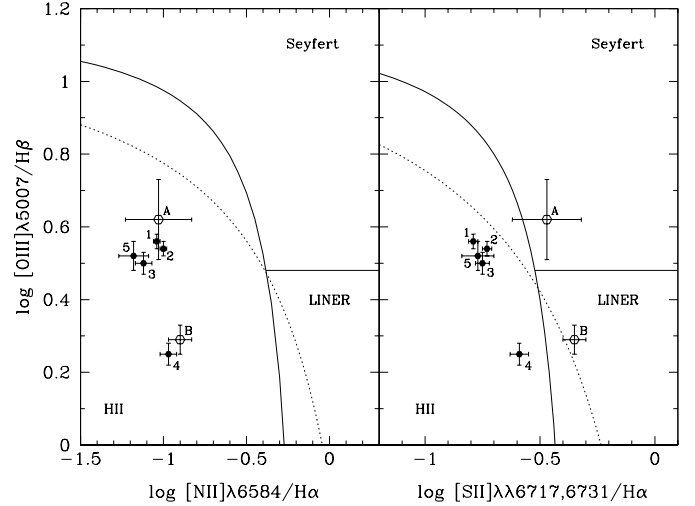}
\caption{Optical emission-line diagnostic diagram for the different SF knots
and the two interknot regions (A and B).
The curves separate Seyfert galaxies, LINERS and \ion{H}{ii} regions-like 
objects; solid lines are the empirical boundaries from \cite{Veilleux1987}, 
while dotted lines are the theoretical boundaries from \cite{Kewley2001}.}
\label{Fig:diagnostic}
\end{figure}

The position of the five SF regions in the diagnostic diagram: 
[\ion{O}{iii}]~$\lambda5007$/\Hb\ vs [\ion{N}{ii}]~$\lambda6584$/\Ha\ and  
[\ion{S}{ii}]~$\lambda\lambda6717,\;6731$/\Ha, is  shown in
Fig.~\ref{Fig:diagnostic}. The  empirical boundaries between the different
zones (from \citealp{Veilleux1987}), as well as the theoretical boundaries
proposed by \cite{Kewley2001}, are also plotted. In both diagrams, all the SF
knots fall within the area occupied by \ion{H}{ii} regions.

\subsection{Kinematics of the ionized gas}
\label{SubSect:Kinematics}

We obtained the velocity field by fitting a single Gaussian to the strong \Ha\
and [\ion{O}{iii}]~$\lambda5007$ emission lines.

\begin{figure*}
\mbox{
\centerline{
\hspace*{0.0cm}\subfigure{\includegraphics[width=6.5cm]{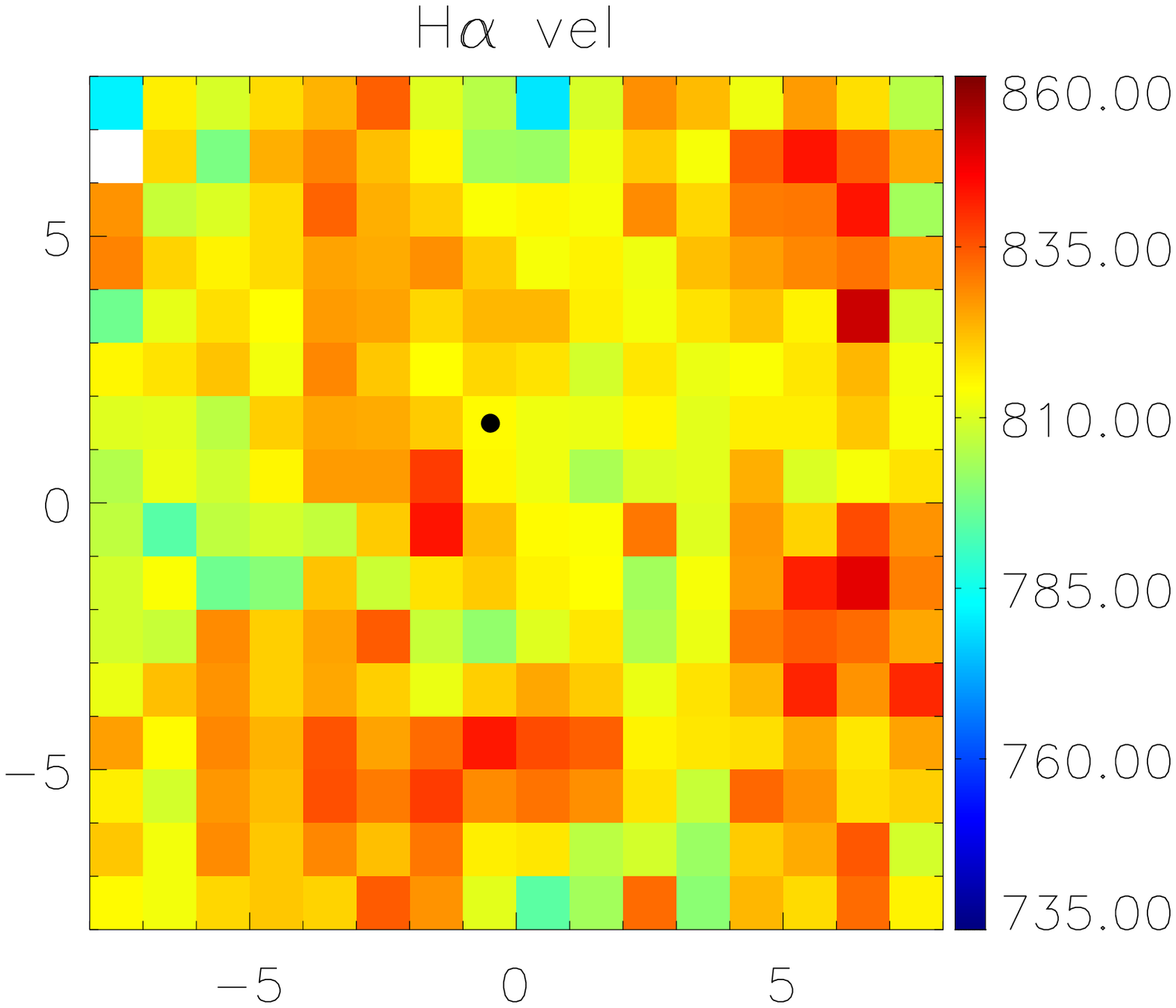}}
\hspace*{0.0cm}\subfigure{\includegraphics[width=6.5cm]{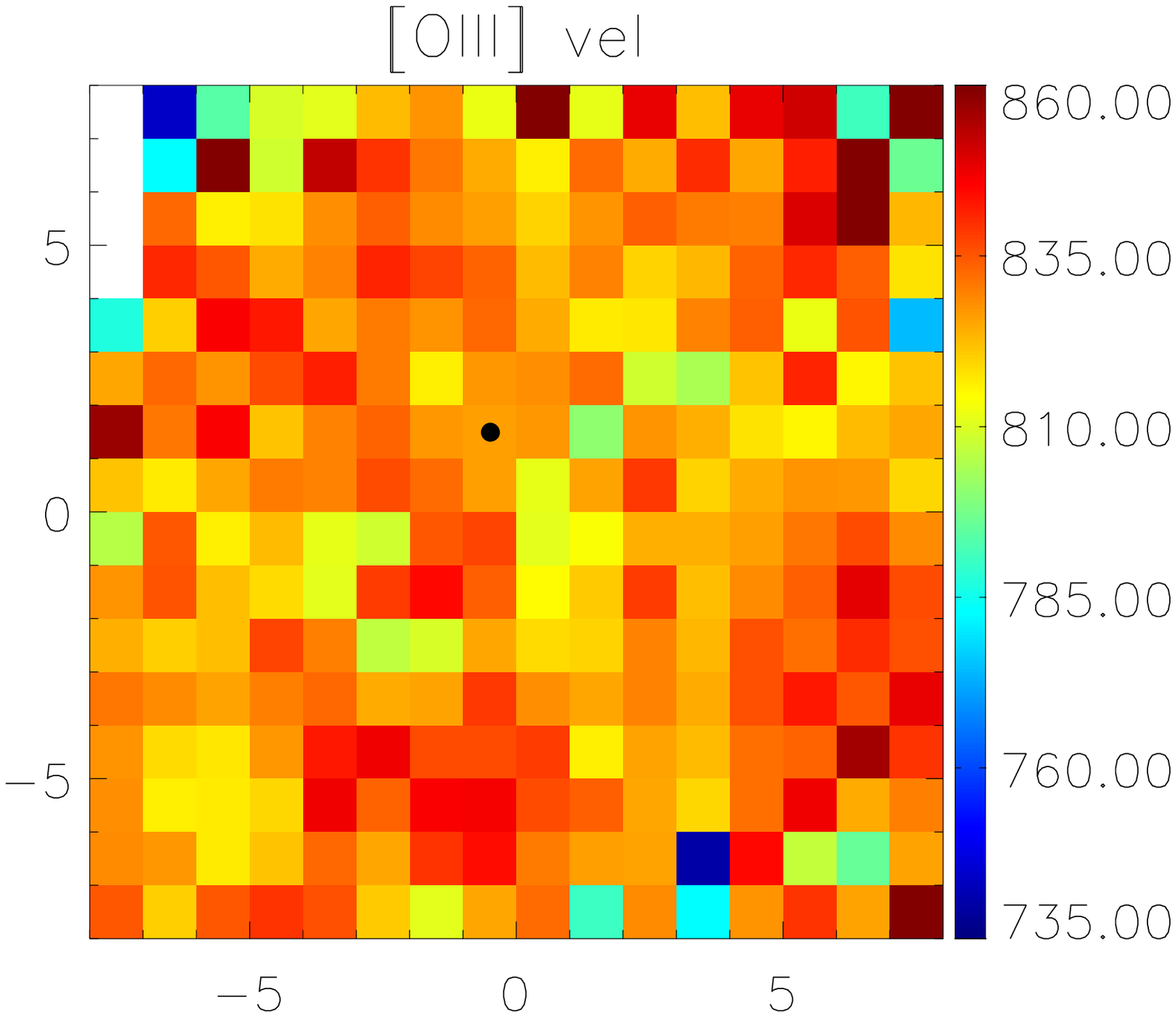}}
}}   
\caption{Velocity field of the ionized gas in the central region of the 
galaxy for the [\ion{O}{iii}]~$\lambda5007$ and the \Ha\ lines. Axis units are 
arcseconds; north is up, east to the left.}
\label{Fig:velocity}
\end{figure*}

Results are displayed in Fig.~\ref{Fig:velocity}. 
No regular pattern is visible in these maps, 
whereas some rotation is expected  in the innermost $10\arcsec$, 
where the velocity changes from 760 to 800 km/s in the radio data
by \cite{VanZee2001} (see their Fig.~12).
Given the spectral resolution of our data, 
$\sim 300$ km s$^{-1}$ at \Ha\ and $\sim 400$ km s$^{-1}$ at \ion{O}{iii},
velocity structures can not be reliably measured. An analysis of the
kinematics of Mrk~1418 in the optical has to await data taken with
higher spectral resolution.

\section{Discussion}
\label{Sect:Discussion}

Results based on PMAS IFS observations of the BCD Mrk~1418 are presented. The
mapped area covers the central starburst region of the galaxy 
($16\times16$ arcsec$^{2}$, equal to $1.14\times1.14$ kpc$^{2}$; see 
Fig.~\ref{Fig:images}).

The emission line and continuum maps reveal quite an intriguing
structure. Both gas and stars display an irregular morphology; interestingly,
none of the peaks in the ionized gas maps coincides with a maximum in the
continuum.
Small spatial offsets between continuum and emission line peaks are
common in compact starburst galaxies, as reported for instance by 
\cite{HunterThronson1995}, \cite{MaizApellaniz1998} and \cite{Lagos2007}, and
are likely related to the release of kinetic energy by massive stars and
supernova explosions.

In the emission lines, at least five major knots are resolved; the two biggest
ones, knot~1 and knot~2, have approximate diameters of 400 pc and 300 pc
respectively. These sizes are comparable to those typical of giant \ion{H}{ii}
regions in M~33 (e.g., NGC~604;\citealp{SandageTammann1974}) and in 
the LMC (e.g., 30~Dor; \citealp{Walborn1991}).

In the continuum maps a bright nucleus is located in the central part
of the field of view, in a place with reduced \Ha\ emission (see
Figure~\ref{Fig:linefluxes}). In the continuum the galaxy shows a distorted,
kidney-shaped morphology, with boxy isocontours, which can be seen with
better clarity in the $B$-band image in Fig.~\ref{Fig:images} and in the
red continuum image in Fig.~\ref{Fig:linefluxes} (see contours). Such boxy
morphology, though usually interpreted as a sign of interactions or mergers
\citep{Doublier1997,Doublier1999}, can just be the result of the superposition
of the starburst on an older, more regular stellar component.

The current SF activity is taking place in the five major line-emitting knots
(Fig.~\ref{Fig:linefluxes}), while the lack of significant gas emission in
the nuclear region suggests the presence of a dominating intermediate age
population. The underlying host galaxy detected in broad-band images
\citep{Doublier1997,GildePaz2003} shows colors ($B-R \sim 1.5-1.7$;
\citealt{Doublier1997,GildePaz2005}) indicative of a population several Gyrs
old.

The ages of the selected regions have been estimated by comparing their \Ha\
and \Hb\ equivalent with the predictions from \textsc{Starburst~99}
\citep{Leitherer1999} evolutionary synthesis models. We chose those models
with $Z=0.4~Z\sun$, the closest value to the metallicity derived
from  the emission-line fluxes ---~a reasonable approximation to the
metallicity of a young population~--- and a Salpeter initial mass function
(IMF) between 1 and 100 $M_{\sun}$.  We found that we can reproduce the
measured \Ha\ and \Hb\ equivalent widths for the five SF knots with an
instantaneous burst of star formation and ages ranging between 5 and 7 Myr:
knots~1 and 2 seem to be the youngest, with equivalent widths consistent with
ages between 5 and 6 Myr, while for knots~3, 4 and 5, we found values slightly
larger, from 6 to 7 Myr. The equivalent widths in the nucleus are consistent
with ages of about 10 Myr.

We should keep in mind that these values are indeed upper limits to the ages
of the knots. In fact, as the continuum flux increases due to the contribution
from the older stars, the measured equivalent widths decrease and this in turn
results into larger derived ages for the knots. As shown in previous
works (\citealt{Cairos2002,Cairos2007}), this effect is not negligible in
BCDs, though difficult to quantify: it rests on the model adopted to describe
the underlying component and can strongly vary with the position of the knots
across the galaxy. For instance, the corrections found for the different knots
in the galaxy Mrk~35 range from 10\% to 90\% \citep{Cairos2007}. By assuming a
contribution of the host galaxy of 50\% to the continuum in the 
brightest knots in  Mrk~1418, the computed age will be about 1 Myr less.

The Interstellar Medium (ISM) around the SF knots is excited by the ionizing
photons coming from the central stars/clusters, with no clear evidence for
other excitation mechanisms. In the ``interknot'' area B a moderate
[\ion{O}{iii}]~$\lambda5007$/\Hb\ ratio together with a large
[\ion{S}{ii}]~$\lambda\lambda6717,\;6731~$/\Ha\ ratio (see 
Fig.~\ref{Fig:diagnostic}) suggest that some
mechanisms other than photoionization contribute to the gas excitation 
(most probably shocks; e.g. \citealp{Shull1979,Schwartz1993}).

On the other hand, nebular \ion{He}{ii}~$\lambda4686$ \AA\ emission has been
detected (in one just spaxel) in a location southwest of the galactic center 
(near knot~2). This line offers a powerful diagnostic of the ionizing fluxes 
and stellar population in starburst regions but, unfortunately, here its  
$S/N$ is too poor for a reliable measurements of its FWHM and flux.

The interstellar extinction has an overall irregular distribution, with a
ridge that crosses the galaxy in a north-south direction between knots~2 and
5, consistent with material swept out by stellar winds and SNe. Its position
and shape are in broad agreement with a dust-lane visible in the SDSS frames.
This finding is in line with recent publications
\citep{Hunt2001,Cairos2003,VanziSauvage2004,Cairos2007} that questioned the
previous belief that BCDs have little or no dust.
\citet{Leroy2005} report Mrk~1418 as a nondetection in their CO 
survey.

In agreement with several other works on BCDs (e.g. \citealp{Izotov1997,
VilchezIglesiasParamo1998, Papaderos2006, Kehrig2008}), we did not find  any
significant chemical abundance variation among the different knots in the
galaxy, which points to a common chemical evolution scenario in all of them.
Also, the nitrogen-to-oxygen abundance ratio values found are consistent with
those obtained for BCDs and \ion{H}{ii} galaxies 
\citep{VilchezIglesiasParamo2003,PerezMontero2005}.

We found for the total spectrum a lower value of direct oxygen abundance and a
higher electron temperature than for knots~1 and 2 (see
Table~\ref{Table:Ratios}).
While this finding may reflect a genuine difference between the oxygen
abundance in the total  spectrum (kpc-sized aperture) and in the
individual \ion{H}{ii} regions (with sizes of the order of 100 pc), it may
also be due to the relatively large measurement uncertainties for the weak 
[\ion{O}{iii}]~$\lambda$4363 emission line.

The origin of the starburst episodes in SF dwarf galaxies is far from being 
well understood. One of the current scenarios is the stochastic 
self-propagating star formation (SSPSF) model proposed by \cite{Gerola1980},
in which the starburst episode in dwarfs is explained as statistical
fluctuations in the SFR. In Mrk~1418 both the ages of the starburst and the
morphology speak against this scenario; the different SF knots have ages too
similar to be causally related: with the typical distances among knots 
(400--800 pc), the star formation will need times of the order of 40--80 Myr 
to propagate through shock traveling at a typical speed of $\sim 10$ km 
s$^{-1}$. 

An alternative scenario, which is gaining popularity, is interactions and
mergers. Recent searches have shown that a substantial fraction of SF dwarfs 
possess low-mass companion galaxies, most of them gas-rich but faint in the
optical \citep{Tayloretal1993,Tayloretal1995,Tayloretal1996,Noeskeetal2001,
Pustilniketal2001}, while studies focused on individual objects have shown
that interactions indeed play a substantial role in the evolution of these
systems
(\citealt{Johnsonetal2004,Ostlin2004,BravoAlfaroetal2004,BravoAlfaroetal2006,
Cumming2008} -- but see \citealt{Vaduvescu2007}).

These observational results have been recently backed up from 
the theoretical point of view. The numerical simulations presented in 
\cite{Bekki2008} can satisfactory explain the physical properties of BCDs 
(e.g. blue compact core, redder low surface brightness component, presence of 
compact young stellar clusters, low metallicity and extended HI gas disk) as the 
result of mergings between dwarfs with larger fraction of gas and extended 
gas disks.

In Mrk~1418, the asymmetric spatial distribution of the SF knots, their
different sizes, the bridge-like structure connecting knot~1 and knot~4, and
the filamentary structures visible in the NED \Ha\ image, as well as the 
boxy, kidney-shape morphology of the continuum, are all suggestive of a system
that has recently experienced some interaction events. Kinematic results from
\ion{H}{i} maps \citep{VanZee2001} also agree with a merger/interaction
scenario: the extended gaseous distribution show kinematical peculiarities,
reminiscent of tidal tails. The \ion{H}{i} velocity field displays a complex
behavior in the inner regions, but an overall clear rotation pattern. 

The low velocity resolution of our PMAS data hampers the derivations of a 
reliable velocity distribution.
Deeper, higher spectral resolution IFU observations, aimed at measuring the 
gas velocity dispersion as well as the stellar kinematics, are essential to 
put tighter constraints on the galaxy dynamics.

\section{Conclusions}
\label{Sect:Conclusions}

The central $16\arcsec\times 16\arcsec$ region of the BCD galaxy Mrk~1418 has
been mapped by using the PMAS spectrograph. Integral Field Spectroscopy
allowed us to simultaneously study the morphology, ionization mechanism,
extinction, chemical abundances and ionized gas kinematics of the galaxy. From
our analysis we highlight the following results:

\begin{enumerate}

\item  Mrk~1418 displays very different morphologies in the continuum and in
the emission lines. While in the continuum the central region has a distorted,
kidney-like shape and boxy isocontours, with the peak located approximately at
the center of the outer isophotes, the emission line maps exhibit five SF
regions irregularly distributed around the optical center.

\item The excitation ratios [\ion{O}{iii}]/\Hb\ and [\ion{N}{ii}]/\Ha\ are
typical of \ion{H}{ii} regions in the whole mapped area, while in the
interknots zones (region B) the [\ion{S}{ii}]/\Ha\ ratios have values
$\geq 0.4$, possibly indicating the presence of shocks.

\item The galaxy displays an inhomogeneous dust distribution, with a peak
value $\Ha/\Hb = 5.5$ in the region between knot~2 and knot~5, about
$3\arcsec$ southeast of the continuum peak; this dust patch has an area of
about $8\sq\arcsec$.

\item For each SF knot we derived line fluxes, interstellar extinction, the
most relevant line ratios and metal abundances. We find that: i) all the
knots show \ion{H}{ii}-like ionization (e.g. star formation); ii) the density
does not vary significantly from knot to knot, and has values typical of giant
extragalactic objects \citep{McCalletal1985}; iii) all knots have similar
metal abundance, with no abundance variations on scales of hundreds of
parsec.

\end{enumerate}

This pilot paper shows that IFS is an essential tool to study BCD galaxies,
as, in just one shot, it allows us to map the gas and stars morphology,
determine the physical parameters and derive chemical abundances. 

Therefore, IFU studies of large BCD samples will enable us to gain insights
into the main issues concerning this class of galaxies, namely, to disentangle
and characterize their stellar populations, to elaborate their SF histories
and to investigate the mechanisms that trigger their star-formation activity.

\begin{acknowledgements} 
L.~M. Cair{\'o}s acknowledges the Alexander von Humboldt Foundation. We thank
J.~N. Gonz{\'a}lez-P{\'e}rez for his help in the initial stages of this
project, A. Monreal-Ibero and B. Garc{\'\i}a-Lorenzo for their assistance with
data reduction and analysis issues, and Y. Ascasibar for fruitful discussions.
N.~C. and C.~Z. are grateful for the hospitality of the Astrophysikalisches
Institut Potsdam. We also thank the anonymous referee for his/her careful and
detailed review of the manuscript. This research has made use of the NASA/IPAC
Extragalactic Database (NED), which is operated by the Jet Propulsion
Laboratory, Caltech, under contract with the National Aeronautics and Space
Administration. This work has been partially funded by the Spanish
``Ministerio de Ciencia e Innovaci{\'o}n'' through grants AYA 2007 67965 and
HA2006-0032, and under the Consolider-Ingenio 2010 Program grant
CSD2006-00070: First Science with the GTC
(http://www.iac.es/consolider-ingenio-gtc/).
\end{acknowledgements}

\bibliographystyle{aa}
\bibliography{ms.bib}

\clearpage 
\onecolumn
\begin{landscape}
\begin{table*}
\begin{scriptsize}
\caption{Reddening corrected line ratios, normalized to \Hb$^{(*)}$, for the 
knots identified in Mrk~1418.}
\label{Table:fluxes_a}
\begin{center}
\begin{tabular}{llcccccccccccccc}       
\hline\hline                 
$\lambda$  & Ion  &         
                     \multicolumn{2}{c}{knot~1}         & 
		     \multicolumn{2}{c}{knot~2}         & 
		     \multicolumn{2}{c}{knot~3}         & 
		     \multicolumn{2}{c}{knot~4}         & 
		     \multicolumn{2}{c}{knot~5}         &  
		     \multicolumn{2}{c}{Nucleus}        & 
		     \multicolumn{2}{c}{Integrated}     \\    
           &      & 
	             F$_{\lambda}$  & -W$_{\lambda}$   & 
		     F$_{\lambda}$  & -W$_{\lambda}$   & 
		     F$_{\lambda}$  & -W$_{\lambda}$   & 
		     F$_{\lambda}$  & -W$_{\lambda}$   & 
                     F$_{\lambda}$  & -W$_{\lambda}$   & 
		     F$_{\lambda}$  & -W$_{\lambda}$   & 
		     F$_{\lambda}$  & -W$_{\lambda}$   \\  
		     \hline   
3727 &  [\ion{O}{ii}]    &  
                    $ 3.19\pm0.27$  &  $ 82.14\pm0.93$  &  
                    $ 3.43\pm0.30$  &  $ 62.70\pm1.03$  &    
                    $ 3.10\pm0.31$  &  $ 52.25\pm1.81$  &   
                    $ 6.00\pm0.56$  &  $117.2 \pm3.53$  &  
                    $ 5.24\pm0.57$  &  $ 65.37\pm3.65$  &  
                    $10.45\pm1.38$  &  $ 23.53\pm0.78$  & 
                    $ 5.44\pm0.43$  &  $ 43.28\pm0.25$  \\
3869 &  [\ion{Ne}{iii}]  &        
                    $ 0.51\pm0.04$  &  $ 11.72\pm0.44$  &  
                    $ 0.39\pm0.03$  &  $  7.94\pm0.44$  &    
                    $ 0.56\pm0.07$  &  $  7.34\pm0.77$  &   
                    $ 0.43\pm0.12$  &  $  6.27\pm1.40$  &  
                    $ 0.83\pm0.20$  &  $  7.68\pm1.62$  & 
                     ---            &  ---              & 
                    $ 1.00\pm0.07$  &  $  7.70\pm0.12$  \\
3968 &  [\ion{Ne}{iii}]  &  
                    $ 0.19\pm0.02$  &  $  4.66\pm0.30$  &  
                    $ 0.18\pm0.02$  &  $  3.74\pm0.36$  &   
                      ---           &  ---              &  
                      ---           &  ---              &  
                      ---           &  ---              &
                     ---            &                   & 
                     ---            &  ---              \\
4101 & H$\delta$ &  
                    $ 0.28\pm0.02$  &  $  6.82\pm0.22$  &    
                    $ 0.30\pm0.02$  &  $  7.31\pm0.32$  &    
                    $ 0.37\pm0.07$  &  $  5.31\pm0.46$  &    
                    $ 0.26\pm0.08$  &  $  3.93\pm0.71$  & 
                      ---           &  ---              &
                      ---           &  ---              & 
                      ---           &  ---              \\
4340 & H$\gamma$ &  
                    $ 0.46\pm0.02$  &  $ 12.95\pm0.23$  &  
                    $ 0.46\pm0.03$  &  $ 13.37\pm0.33$  &    
                    $ 0.46\pm0.04$  &  $  7.21\pm0.36$  &    
                    $ 0.47\pm0.05$  &  $  8.28\pm0.60$  &  
                    $ 0.47\pm0.07$  &  $  6.91\pm0.63$  & 
		    --- 	    &  ---              & 
		    $ 0.47\pm0.02$  &  $  4.29\pm0.05$	\\
4363 & [\ion{O}{iii}]    &  
                    $ 0.05\pm0.01$  &  $  2.00\pm0.23$  &  
                    $ 0.05\pm0.01$  &  $  2.29\pm0.37$  &    
                     ---            &  ---              &    
                     ---            &  ---              &  
                     ---            &  ---              &
                    ---             &  ---              & 
                    $ 0.10\pm0.04$  &  $ 1.02 \pm0.06$  \\
4861 & H$\beta$          &  
                    $ 1.00       $  &  $ 31.27\pm0.25$  &  
                    $ 1.00       $  &  $ 32.96\pm0.34$  &    
                    $ 1.00       $  &  $ 17.42\pm0.39$  &    
                    $ 1.00       $  &  $ 18.30\pm0.90$  &  
                    $ 1.00       $  &  $ 12.73\pm0.68$  &
                    $ 1.00       $  &  $  3.03\pm0.24$  & 
                    $ 1.00       $  &  $ 10.14\pm0.06$  \\
4959 & [\ion{O}{iii}]   &  
                    $ 1.24\pm0.04$  &  $ 38.81\pm0.29$  &  
                    $ 1.16\pm0.03$  &  $ 41.07\pm0.44$  &    
                    $ 1.12\pm0.05$  &  $ 19.89\pm0.41$  &    
                    $ 0.61\pm0.04$  &  $ 11.54\pm0.48$  &  
                    $ 1.14\pm0.08$  &  $ 15.78\pm0.66$  &
                    $ 0.83\pm0.09$  &  $  2.61\pm0.18$  & 
                    $ 1.06\pm0.03$  &  $ 11.01\pm0.06$  \\
5007 &  [\ion{O}{iii}]   &  
                    $ 3.62\pm0.10$  &  $113.9 \pm0.46$  &  
                    $ 3.45\pm0.10$  &  $122.70\pm0.51$  &    
                    $ 3.19\pm0.12$  &  $ 55.06\pm0.49$  &     
                    $ 1.78\pm0.09$  &  $ 34.54\pm0.54$  &  
                    $ 3.29\pm0.18$  &  $ 46.81\pm0.90$  &
                    $ 2.34\pm0.20$  &  $  7.61\pm0.20$  & 
                    $ 3.16\pm0.09$  &  $ 33.48\pm0.09$  \\
5876 &  \ion{He}{i}      &  
                     ---            &   ---             &  
                    $ 0.07\pm0.01$  &  $  3.11\pm0.21$  &    
                    $ 0.11\pm0.01$  &  $  2.48\pm0.26$  &    
                    $ 0.07\pm0.01$  &  $  1.88\pm0.35$  &  
                    $ 0.08\pm0.02$  &  $  1.60\pm0.49$  &  
                    ---             &   ---             &  
                    $ 0.05\pm0.01$  &  $  0.75\pm0.03$  \\
6548 &  [\ion{N}{ii}]    &  
                    $ 0.08\pm0.01$  &  $  3.69\pm0.19$  &  
                    $ 0.15\pm0.01$  &  $  8.15\pm0.44$  &    
                    $ 0.10\pm0.02$  &  $  2.65\pm0.34$  &   
                    $ 0.13\pm0.02$  &  $  3.77\pm0.47$  &  
                    $ 0.10\pm0.03$  &  $  2.37\pm0.62$  &  
                    $ 0.12\pm0.03$  &  $  0.80\pm0.21$  & 
                    $ 0.11\pm0.01$  &  $  1.95\pm0.05$  \\
6563 & H$\alpha$ &  
                    $ 2.87\pm0.10$  &  $137.6 \pm0.39$  &  
                    $ 2.88\pm0.14$  &  $154.1 \pm1.35$  &    
                    $ 2.90\pm0.20$  &  $ 73.72\pm0.79$  &    
                    $ 2.87\pm0.15$  &  $ 84.12\pm1.41$  &  
                    $ 2.87\pm0.16$  &  $ 69.03\pm1.48$  &
                    $ 2.86\pm0.34$  &  $ 19.63\pm0.27$  & 
                    $ 2.86\pm0.07$  &  $ 49.87\pm0.10$  \\ 
6584 &  [\ion{N}{ii}]    &  
                    $ 0.26\pm0.01$  &  $ 12.50\pm0.23$  &  
                    $ 0.29\pm0.02$  &  $ 15.22\pm0.43$  &    
                    $ 0.22\pm0.02$  &  $  5.68\pm0.40$  &   
                    $ 0.31\pm0.03$  &  $  9.14\pm0.61$  &  
                    $ 0.19\pm0.03$  &  $  4.56\pm0.61$  &
                    $ 0.33\pm0.05$  &  $  2.25\pm0.20$  & 
                    $ 0.27\pm0.01$  &  $  4.63\pm0.06$  \\
6717 & [\ion{S}{ii}]     &  
                    $ 0.26\pm0.01$  &  $ 12.50\pm0.28$  &  
                    $ 0.31\pm0.02$  &  $ 17.05\pm0.44$  &    
                    $ 0.31\pm0.03$  &  $  8.21\pm0.44$  &   
                    $ 0.42\pm0.03$  &  $ 12.93\pm0.60$  &  
                    $ 0.29\pm0.03$  &  $  7.30\pm0.66$  &
                    $ 0.58\pm0.08$  &  $  4.21\pm0.23$  &
                    $ 0.40\pm0.01$  &  $  8.49\pm0.05$  \\ 
6731 & [\ion{S}{ii}]          &  
                    $ 0.21\pm0.01$  &  $ 10.22\pm0.25$  &  
                    $ 0.23\pm0.02$  &  $ 12.69\pm0.43$  &    
                    $ 0.20\pm0.02$  &  $  5.43\pm0.41$  &   
                    $ 0.32\pm0.02$  &  $  9.87\pm0.57$  &  
                    $ 0.20\pm0.03$  &  $  5.17\pm0.68$  &
                    $ 0.42\pm0.06$  &  $  3.02\pm0.21$  &
                    $ 0.30\pm0.01$  &  $  5.52\pm0.07$  \\ 
      &          &  
                    \multicolumn{2}{c}{$F(\Hb)= 324.6\pm22.7$} &
                    \multicolumn{2}{c}{$F(\Hb)= 166.3\pm15.2$} &
                    \multicolumn{2}{c}{$F(\Hb)=  67.4\pm11.3$} &
                    \multicolumn{2}{c}{$F(\Hb)=  84.6\pm 6.0$} &
                    \multicolumn{2}{c}{$F(\Hb)=  42.4\pm 3.9$} &
                    \multicolumn{2}{c}{$F(\Hb)= 106.3\pm30.4$} &
                    \multicolumn{2}{c}{$F(\Hb)=1871.9\pm62.6$} \\
      &          &              
		    \multicolumn{2}{c}{$C(\Hb)=0.111\pm0.032$} &
                    \multicolumn{2}{c}{$C(\Hb)=0.073\pm0.035$} &
                    \multicolumn{2}{c}{$C(\Hb)=0.123\pm0.064$} &
                    \multicolumn{2}{c}{$C(\Hb)=0.267\pm0.025$} &
                    \multicolumn{2}{c}{$C(\Hb)=0.361\pm0.030$} &
                    \multicolumn{2}{c}{$C(\Hb)=0.625\pm0.105$} &
                    \multicolumn{2}{c}{$C(\Hb)=0.316\pm0.011$} \\
      &          &	    
                    \multicolumn{2}{c}{$W_\mathrm{abs}=3.2\AA$} &
                    \multicolumn{2}{c}{$W_\mathrm{abs}=2.9\AA$} &
                    \multicolumn{2}{c}{$W_\mathrm{abs}=2.9\AA$} &
                    \multicolumn{2}{c}{$W_\mathrm{abs}=1.7\AA$} &
                    \multicolumn{2}{c}{$W_\mathrm{abs}=1.4\AA$} &
                    \multicolumn{2}{c}{$W_\mathrm{abs}=3.0\AA$} &
                    \multicolumn{2}{c}{$W_\mathrm{abs}=2.1\AA$} \\
\hline
\end{tabular}
\end{center}
(*) Notes. -- Reddening-corrected line fluxes, normalized to $F(\Hb)=1$. 
Equivalent widths of Balmer lines are corrected for underlying stellar 
absorption. The reddening coefficients, $C(\Hb)$, the value of the absorption 
correction, EW$_\mathrm{abs}$, and the reddening-corrected \Hb\ flux, $F(\Hb) 
(\times 10^{-16}$ ergs cm$^{-2}$ s$^{-1}$) are listed for each region. 
In the nuclear region, only \Hb\ and \Hb\ could be measured, 
so we were unable to compute EW$_\mathrm{abs}$ and $C(\Hb)$ simultaneously; 
to determine the latter we assumed EW$_\mathrm{abs}=3.0$. 
The quoted uncertainties include measurements, flux-calibration and 
reddening coefficient errors. 
\end{scriptsize}
\end{table*}
\end{landscape}

\begin{landscape}
\begin{table*}
\caption{Physical parameters and abundances}
\label{Table:Ratios}
\begin{center}
\begin{tabular}{lccccccc} 
\hline\hline  
 Parameter  &  knot~1   & knot~2 &  knot~3 & knot~4 & knot~5 & Nucleus & 
               Integrated \\ 
\hline   
$\log\left(\frac{[\mbox{\ion{O}{iii}}]\;\lambda5007}{\Hb}\right)$       & 
        $ 0.56\pm0.02$  & 
        $ 0.54\pm0.02$  & 
        $ 0.50\pm0.03$  &
        $ 0.25\pm0.03$  & 
        $ 0.52\pm0.04$  & 
        $ 0.37\pm0.05$  & 
        $ 0.50\pm0.02$  \\
$\log\left(\frac{[\mbox{\ion{N}{ii}}]\;\lambda6584}{\Ha}\right)$       & 
        $-1.04\pm0.02$  & 
        $-1.00\pm0.02$  & 
        $-1.12\pm0.05$  &   
	$-0.97\pm0.05$  & 
        $-1.18\pm0.09$  & 
        $-0.94\pm0.08$  & 
        $-1.03\pm0.02$  \\
$\log\left(\frac{[\mbox{\ion{S}{ii}}]\;\lambda\lambda6717,\;6731}{\Ha}\right)$    & 
        $-0.79\pm0.02$  & 
        $-0.73\pm0.02$  & 
        $-0.75\pm0.03$  &   
	$-0.59\pm0.04$  & 
        $-0.77\pm0.07$  & 
        $-0.46\pm0.06$  & 
        $-0.61\pm0.02$  \\\hline
$N_\mathrm{e}$([\ion{S}{ii}]) (cm$^{-3}$)       & 
        190             & 
        $\leq 100$      &   
        $\leq 100$      &   
	108             &  
        $\leq100$       &  
        $\leq100$       & 
        $\leq100$       \\
$T_{\mathrm e}$([\ion{O}{ii}]) (10$^{4}$ K)        & 
        $1.18\pm0.07$   & 
        $1.22\pm0.07$   & 
        ---             & 
	---             & 
        ---             &  
        ---             & 
        $1.67\pm0.30$   \\ 
$T_{\mathrm e}$([\ion{O}{iii}]) ($10^{4}$ K)       & 
        $1.27\pm0.10 $  & 
        $1.33\pm0.10 $  & 
        ---             & 
	---             & 
        ---             &  
        ---             & 
        $1.96\pm0.42$   \\ 
$12+\log(O/H)$ -- ($T_{e}$)         & 
        $8.09\pm0.06 $  & 
        $8.03\pm0.06 $  & 
        ---             & 
	---             & 
        ---             & 
        ---             & 
       $7.71\pm0.13$   \\
$12+\log(\mathrm {O/H})$ -- ($N2$)                                             & 
        8.26            & 
        8.28            & 
        8.23            & 
	8.30            & 
        8.20            & 
        8.31            & 
        8.27            \\
$12+\log(\mathrm{O/H})$ -- ($O3N2$)                                            & 
        8.22            & 
        8.24            & 
        8.21            & 
	8.34            & 
        8.19            & 
        8.31            & 
        8.24            \\
$12+\log(\mathrm{S}^{+}/\mathrm{H}^{+})$                &
        $5.87\pm0.05$   & 
    	$5.90\pm0.05$   &
    	---             &
    	---             &
    	---             &
    	---             &
    	$5.80\pm0.10$   \\
$12+\log(\mathrm{N}^+/\mathrm{H}^+)$         &
        $6.52\pm0.05$   &
	$6.59\pm0.05$   &
        ---             &
	---             &
	---             &
	---             &
	$6.26\pm0.11$   \\
$\log(\mathrm{N/O})$             &
       $-1.29\pm0.10$   &
       $-1.18\pm0.10$   &
        ---             &
        ---             &
        ---             &
	---             &
       $-1.27\pm0.21$   \\
\hline
\end{tabular}
\end{center}

Notes:
$T_\mathrm{e}$ ([\ion{O}{ii}]) derived from the relation: 
$T_\mathrm{e} \mathrm{([\ion{O}{ii}])}=0.72\times T_{\mathrm e} 
\mathrm{([\ion{O}{iii}])} + 0.26$ found by \cite{Pilyugin2006}. \\
$T_\mathrm{e}$ ([\ion{O}{iii}]): electron temperature measured from 
[\ion{O}{iii}]~$\lambda4363$  \\
$12+\log(\mathrm{O/H})$ -- ($T_\mathrm{e}$): direct O/H abundance derived from 
$T_\mathrm{e}$[\ion{O}{iii}] \\  
$12+\log(\mathrm{O/H})$ -- ($N2$): O/H derived from the $N2$ index 
\citep{PettinPagel2004}; the associated uncertainty is $\pm0.38$  \\
$12+\log(\mathrm{O/H})$ -- ($O3N2$): O/H derived from the 
$O3N2$ index \citep{PettinPagel2004}; the associated 
uncertainty is $\pm0.25$. 
\end{table*}
\end{landscape}

\end{document}